\documentclass[12pt]{article}
\usepackage{graphicx, epsfig}
\usepackage{amsmath}
\usepackage{latexsym}
\usepackage{amstext}
\usepackage{amssymb}
\newcommand{\mb}[1]{ \mbox{\boldmath$#1$} }
\newcommand{\vect}[1]{\boldsymbol{#1}}
\newcommand{\ds}{\displaystyle}
\newcommand{\beq}{\begin{eqnarray}}
\newcommand{\eeq}{\end{eqnarray}}
\newcommand{\beqq}{\begin{eqnarray*}}
\newcommand{\eeqq}{\end{eqnarray*}}
\newcommand{\p}{\partial}

\newcommand{\eps}{\varepsilon}
\newcommand{\x}{\mbox{\boldmath$x$}}

\newcommand{\AB}{\mbox{\boldmath$AB$}}
\newcommand{\y}{\mbox{\boldmath$y$}}
\newcommand{\nn}{\nonumber}

\newcommand{\n}{\mbox{\boldmath$n$}}

\newcommand{\w}{\mbox{\boldmath$w$}}

\renewcommand{\(}{\left(}
\renewcommand{\)}{\right)}
\font\bb=msbm10 at 12pt
\def\rR{\hbox{\bb R}}

\def\eE{\hbox{\bb E}}

\begin{document}

%\pagestyle{plain}
%%%%%%%%%%%%%%%%%%%%%%%%%%%%%%%%%%%%%%%%%%%%%%%%%%%%%%%%%%%%%%%%%%%%%%%%%%%%
\begin{center}
{\large \textbf{100 years after Smoluchowski: stochastic processes in cell biology}}\\[5mm]
D. Holcman$^{*}$ \footnote{Applied Mathematics and Computational Biology, Ecole Normale Sup\'erieure, France and Churchill College, DAMTP Cambridge university, CB3 0DS, UK. $^{*}$corresponding authors: david.holcman@ens.fr},  and Z. Schuss \footnote{Department of Applied Mathematics, Tel-Aviv University, Tel-Aviv 69968, Israel.}
\end{center}
%Eigenvalues of the noise-activated escape over a limit cycle
\date{}
\begin{abstract}
100 years after Smoluchowski introduces his approach to stochastic processes, they are now at the basis of mathematical and physical modeling in cellular
biology: they are used for example to analyse and to extract features from large number (tens of thousands) of single molecular trajectories or to study the diffusive motion of molecules, proteins or receptors.  Stochastic modeling is a new step in large data analysis that serves extracting cell biology concepts. We review here the Smoluchowski's approach to stochastic processes and provide several applications for coarse-graining diffusion, studying polymer models for understanding nuclear organization and finally, we discuss the stochastic jump dynamics of telomeres across cell division and stochastic gene regulation.
\end{abstract}

\tableofcontents \markboth{Contents}{Contents}

%%%%%%%%%%%%%%%%%%%%%%%%%%%%%%%%%%%%%%%%%%5
\section{Introduction}
%%%%%%%%%%%%%%%%%%%%%%%%%%%%%%%%%%%%%%%%%%5
Stochastic processes have become the cornerstone of mathematical and physical modeling in cellular
biology.  Their role is to simulate and predict measurable cellular biology phenomena from molecular level physics. The early theory began with Einstein's early work on the Brownian motion in configuration space \cite{Einstein}, which was extended by Langevin \cite{Langevin} to the stochastic description of Brownian motion in phase space. A major advancement was Smoluchowski's observation \cite{Smoluchowski} that in the overdamped regime of Langevin's phase space-model, displacement and velocity become statistically independent and thus the analysis of the Langevin model reduces to that of displacement in configuration space alone, while velocities remain Maxwellian. This reduction made the use of the Fokker-Planck-Smoluchowski-Kolmogorov partial differential equations the main tool for extracting probabilistic and thus physical and chemical information from molecular models. In particular, the solution of boundary value problems for these partial differential equations clarified the significance of the mean escape time of a Smoluchowski configuration-space trajectory from the domain of attraction of a stable attractor, such as the escape over a potential barrier, called thermal activation. Chandrasekhar's 1943 paper \cite{Chandrasekhar} reviews the early period of stochastic models in different branches of physics, chemistry and astronomy. In particular, it reviews Kramers' theory \cite{Kramers} of thermal activation over a potential barrier and his development of an asymptotic method for the approximation of the solution of singular perturbation problems for the Fokker-Planck-Smoluchowski partial differential equation. Kramers' theory was developed further in the 1950s-2000s and applied in many physical, chemical, and engineering problems, such as impurity diffusion in crystals, the transitions between super-conducting and conducting states of the driven Josephson junction, loss of lock in tracking loops, and other non-equilibrium processes \cite{Schuss1,Schuss2}.

In the early 1970s a new direction has emerged. Merton, Black, and Scholes developed a Smoluchowski-type stochastic model for the market value of stock and used it to predict the value of a future contract on the stock, given its current market value (stock option pricing theory), for which they were awarded the 1997 Nobel Prize in economics. Since the publication of their paper in 1973 the market of stock derivatives exploded and the value of traded derivatives became significantly bigger than all  commodities put together, including real estate in downtown Tokyo.

Following the work of Einstein, Langevin, and Smoluchowski a partial differential equation was derived by Fokker in 1914 (for a linear model) and in 1917 by Planck for the general Langevin equation (see \cite{Chandrasekhar,Schuss2} for references). The rigorous mathematical treatment of the relationship between Langevin trajectories and partial differential equations began in the 1930s with the formalism of Kolmogorov and Wiener, who put it into abstract probability theory. In the mathematical literature it became clear that in addition to It\^o's formulation, which assumed
that the noise and the state are independent, there is another formulation, due to Stratonovich, that takes into account correlations between the state and the noise. This is important both in modeling and simulations. The question which form, It\^o or Stratonovich, is correct for the given model, is answered at the microscopic level, not at the equation level (see discussion in \cite{Schuss2}).

The calculation of the mean first passage time (MFPT) for noise-activated escape from the an attractor leads to singular perturbation problems in parabolic and elliptic partial differential equations. Therefore analytical approximations of the solution require asymptotic methods that were developed in fluid dynamics \cite{VanDyke} and quantum mechanics \cite{Wasow,Bender}, such as boundary layer theory \cite{Kevorkian,Bender}, matched asymptotics \cite{Lagestrom}, the WKB method \cite{Bender}, \cite{Kevorkian,Olver}, and more.

Molecular and cellular biophysics were relatively late comers to the world of stochastics. They introduced a slew of new mathematical problems in stochastics. Thus, for instance, Smoluchowski and Smoluchowski-Stratonovich equations appear as approximations to Markovian models in the continuum limit. Deviations from the Smoluchowski-Stratonovich model are expressed by the introduction of a memory kernel, which may represent a coarse-grained model of interaction of the molecular path with many degrees of freedom, such as the medium in which a molecule is immersed, coupling between molecules, and so on  \cite{Schuss2}.

In the 1980s and 1990s, motivated by biophysical questions, modeling and simulations were developed to simulate and analyze the motion
of ions inside protein channels of biological membranes or across different concentrations \cite{Hille}, \cite{Eisenberg}. Later, in the 1990s and the 2000s, a new field emerged that used Smoluchowski dynamics to predict biological processes on the molecular level. Specifically, the accumulation of massive physiological data prompted the use of stochastic models of synaptic transmission in neurobiology, of calcium  dynamics in microdomains, the motion of ions in selective ion channels, and so. The narrow escape theory emerged as a generic theory to study the rare cellular events of arrival of Brownian particles at a small absorbing part of an impermeable membrane, which may represent any small target for diffusing molecules \cite{HolcmanSchuss2015,Ward1,Ward2,Dimarzo,HS1,PNAS1,NarrowEscape1,NarrowEscape2,NarrowEscape3,Coombs,Cheviakov}.

Nowadays, 100 years after Smoluchowski, the field of stochastic modeling has matured into molecular biophysics and physiology and brought with it a plethora of new mathematical problems. A particularly fruitful direction is that of calculating the MFPT of a Smoluchowski trajectory to a small target, the so-called narrow escape problem. Obviously, the Smoluchowski trajectory may represent that of a molecule inside or outside a biological cell or on its membrane. The target is small in the sense that its size is much smaller than that of the cell. Although similar to the activation problem, the narrow escape problem is radically different and calls for new asymptotic methods. Curiously enough, the same mathematical problem appears in Helmholtz's theory of radiation through a small opening \cite{Helmholtz}. The new narrow escape theory (NET), which appeared initially as a new mathematical first passage time problem, is the
centerpiece of this review.

%
%
%This review-survey discusses the following topics:
%%%%%%%%%%%%%%%%%%%%%%%%%%%%%%%%%%%%%%%%%%%
%\subsection{Organization of the review}
%%%%%%%%%%%%%%%%%%%%%%%%%%%%%%%%%%%%%%%%%%%
%\begin{itemize}
%\item The stochastic equation: framework.
%\item Narrow escape theory in cell biology: {Chapter 4 of the Narrow Escape book}
%\item Super-resolution analysis: { Chapter 5 of the Narrow Escape book and the paper \textit{ Review-short-trajectories2}}
%\item Simulations of stochastic processes in empirical domains and stochastic simulations {in shaped domains and of a needle: Section 5.2 of the Narrow Escape book and sections 2.6 and 7.7 (diffusion of a needle) in my book \textit{Brownian Dynamics at Boundaries and Interfaces.}} {Do we have cell biology applications for shaped objects?}
%\item Facilitated diffusion and gene regulation: continuum of Krox 20 expression.
%\item Polymeric processes: looping, translocation
%\end{itemize}

%%%%%%%%%%%%%%%%%%%%%%%%%%%%%%%%%%%%%%%%%%
%\section{From Einstein to Smoluchowski}
%%%%%%%%%%%%%%%%%%%%%%%%%%%%%%%%%%%%%%%%%%
%%%%%%%%%%%%%%%%%%%%%%%%%%%%%%%%%%%%%%%%%%
\section{Construction of the Brownian trajectories}
%%%%%%%%%%%%%%%%%%%%%%%%%%%%%%%%%%%%%%%%%%
The laws of diffusion were first formulated by Fick. His first law
of diffusion, formulated in 1856 by analogy with Fourier's first law
of heat conduction, asserts that \textit{the diffusion flux between two
points of different concentrations in the fluid is proportional to
the concentration gradient between these points}. The constant of
proportionality is called \textit{the diffusion
coefficient\index{diffusion coefficient}} and it is measured in
units of area per unit time.

In 1905 Einstein\index{Einstein, A.} \cite{Einstein}\index{Einstein,
A.} and, independently, in 1906 Smoluchowski\index{Smoluchowski, M.
von} \cite{Smoluchowski} offered an explanation of the Brownian
motion\index{Brownian motion} based on kinetic theory and
demonstrated, theoretically, that the phenomenon of diffusion is the
result of Brownian motion\index{Brownian motion}.
Einstein's\index{Einstein, A.} theory was later verified
experimentally by Perrin \cite {Perrin}\index{Perrin, J.} and
Svedberg \cite{Svedberg1}\index{Svedberg, T.}. That of
Smoluchowski\index{Smoluchowski, M. von} was verified by
Smoluchowski\index{Smoluchowski, M. von}
\cite{S1}\index{Smoluchowski, M. von}, Svedberg
\cite{Svedberg2} and Westgren \cite{Westgren}, \cite{Westgren1}. Perrin has won the 1926 Physics Nobel Prize for his experiment.

They derived an explicit formula for the diffusion coefficient,
\begin{align}
D=\frac{RT}{N}\frac{1}{6\pi a\eta},  \label{DC}
\end{align}
where $R$ is the universal gas constant, $T=$ absolute (Kelvin) temperature,
and $N=$ Avogadro's number, $a=$ radius of the particle, and $\eta=$ coefficient of
dynamical viscosity.
Equation (\ref{DC}) was obtained from similar considerations by
Sutherland in 1904 and published in 1905 \cite{Sutherland}, but has never received due credit for it.

To connect this theory with the ``\textit{irregular movement which arises
from thermal molecular movement,}'' Einstein\index{Einstein, A.} made
the following assumptions: \textit{(1) the motion of each particle is
independent of the others and (2) ``the movements of one and the
same particle after different intervals of time must be considered
as mutually independent processes, so long as we think of these
intervals of time as being chosen not too small.}''  He derived from these assumptions the diffusion equation for the density
function $p(x,t)$ of finding the Brownian particle at point $x$ on the line at time $t$ and its solution
\begin{align}
p(x,t)=\frac{1}{\sqrt{4\pi Dt}}
\exp\!\left\{-\frac{x^2}{4Dt}\right\}, \label{pdf}
\end{align}
which can be interpreted as the transition probability
density of a particle from the point $x=0$ at time $0$ to the point $x$ at time $t$.

If we denote by $x(t)$ the displacement (or trajectory) of the particle at time $t$,
then for any spatial interval $A$,
\begin{align}
\Pr\left\{x(t)\in A\right\}=\int\limits_Ap\,(x,t)\,dx.  \label{Pr}
\end{align}
It follows that the moments of the Brownian displacement process are
\begin{align}
\eE x(t)=\int xp\,(x,t)\,dx=\eE x(t)=0,\quad\eE x^2(t)=2Dt. \label{moments}
\end{align}
Obviously, if the particle starts at $x(0)=x_0$, then
\begin{align}
\eE[x(t)\,|\,x(0)=x_0]=&\,x_0\label{moments0}\\
\mbox{Var}[x(t)\,|\,x(0)=x_0]=&\,\eE[(x(t)-x_0)^2\,|\,x(0)=x_0]=
2Dt.\nonumber
\end{align}

Now, using eq. (\ref{DC}) in eq. (\ref{moments0}), the mean square
displacement of a Brownian particle along the $x$-axis is found as
\begin{align}
\sigma=\sqrt{t}\sqrt{\frac{kT}{3\pi a\eta}},  \label{sigam}
\end{align}
where $k=R/N$ is Boltzmann's constant. This formula was verified
experimentally \cite{Svedberg1}\index{Svedberg, T.}. It indicates
that the mean square displacement of a Brownian particle at times
$t$ not too short is proportional to the square root of time.

The mathematical question of the existence of a stochastic (random) process which satisfies Einstein's requirements and of its actual construction, was answered in the affirmative in 1933 by Paley, Wiener, and Zygmund \cite{PWZ}, who constructed the random Brownian
trajectories in the form of a Fourier series with random Gaussian coefficients. They proved, \textit{i.a.}, that the Brownian trajectories are nowhere differentiable with probability 1. Another, more modern approach, was proposed by P. L\'evy \cite{Levy}.
L\'evy's construction of a Brownian path in the time interval $[0,1]$ consist in refining linear interpolations of points sampled independently from the Gaussian distribution at binary times $t_{k,n}=k2^{-n},\,(k=0,1\ldots,n)$, such that the properties \eqref{moments}, \eqref{moments0} are satisfied at the binary times $t_{k,n}$ (see details in \cite{Schuss2}).
%\subsection{When Langevin equation is necessary: channels, get velocity}
\section{The velocity process and Langevin's approach\label{s:velocity}}

\subsection{The velocity problem}Obviously, the infinite velocities of the Brownian trajectories contradict physics. Specifically, according to the Waterston-Maxwell equipartition theorem \cite{Waterson}, the root
mean square (RMS) velocity $\bar v=\sqrt{\eE v^2}$ of a
suspended particle should be determined by the equation
\begin{align}
\frac{m}{2}{\bar v^2}=\frac{3kT}{2}.  \label{WM}
\end{align}
Each component of the velocity vector has the same variance, so that
\begin{align}
\frac{m}{2}{\bar v_{x,y,z}^2}=\frac{kT}{2},  \label{WM1d}
\end{align}
which is the one-dimensional version of eq. (\ref{WM}). The RMS
velocity comes out to be about 8.6\,cm/sec for the particles used in
Svedberg's experiment \cite{Svedberg1}\index{Svedberg, T.}.
Einstein\index{Einstein, A.} argued in 1907 and 1908
\cite{Einstein}\index{Einstein, A.} that there is no possibility of
observing this velocity, because of the very rapid viscous damping,
which can be calculated from the Stokes formula. The velocity of
such a particle would drop to 1/10 of its initial value in about
$3.3\times10^{-7}\,$sec. Therefore, Einstein\index{Einstein, A.}
argued, in the period $\tau$ between observations the particle must
get new impulses to movement by some process that is the inverse of
viscosity, so that it retains a velocity whose RMS average is $\bar
v$. Between consecutive observations these impulses alter the
magnitude and direction of the velocity in an irregular manner, even
in the extraordinarily short time of $3.3\times10^{-7}\,$sec.
According to this theory, the RMS velocity in the interval $\tau$
has to be inversely proportional to $\sqrt{\tau}$; that is, it
increases without limit as the time interval between observations
becomes smaller.

\subsection{Langevin's solution of the velocity problem}In 1908 Langevin\index{Langevin's equation}
\cite{Langevin}\index{Langevin, P.} offered an alternative approach
to the problem of the Brownian motion\index{Brownian motion}. He
assumed that the dynamics of a free Brownian particle is governed by the frictional force $-6\pi a\eta
v$ and by a fluctuational force $\Xi$ that results from the random
collisions of the Brownian particle with
the molecules of the surrounding fluid, after the frictional force
is subtracted. This force is random and assumes positive and
negative values with equal probabilities. It follows that Newton's
second law of motion for the Brownian particle is given by
\begin{align}
m\ddot x=-6\pi a\eta\dot x+\Xi.  \label{LE1}
\end{align}
Denoting $v=\dot x$ and multiplying eq. (\ref{LE1}) by $x$, we
obtain
\begin{align}
\frac{m}{2}\frac{d^2}{dt^2}x^2-mv^2=-3\pi a\eta\frac{d}{dt}x^2+\Xi
x. \label{Xx}
\end{align}
Averaging under the assumption that the fluctuational force $\Xi$
and the displacement of the particle $x$ are independent, we obtain
\begin{align}
\frac{m}{2}\frac{d^2}{dt^2}\eE x^2+3\pi
a\eta\frac{d}{dt}\eE x^2=kT,  \label{ALE}
\end{align}
where \eqref{WM1d} has been used. The solution is given by
$d\eE x^2/dt=kT/3\pi a\eta+Ce^{-6\pi a\eta t/m}$, where
$C$ is a constant. The time constant in the exponent is $10^{-8}\,$
sec, so the mean square speed decays on a time scale much shorter
than that of observations. It follows that $\eE x^2-\eE x^2_0=(kT/3\pi a\eta)t$. This, in turn
(see \eqref{moments}), implies that the diffusion
coefficient is given by $D=kT/6\pi a\eta$, as in Einstein's equation \eqref{DC}.

Langevin's equation(\ref{LE1}) is a \textit{stochastic differential equation\index{stochastic differential
equation}}, because it is driven by a random force\index{random
force} $\Xi$. If additional fields of force act on the diffusing
particles (e.g., electrostatic, magnetic, gravitational, etc.),
Langevin's equation is modified to
include the external force, $F(x,t)$, say,
\cite{Kramers}\index{Kramers, H.A.},
\cite{Chandrasekhar},
\begin{align}
m\ddot x+\Gamma \dot x-F(x,t)=\Xi,  \label{FLE}
\end{align}
where $\Gamma= 6\pi a\eta$ is the friction coefficient of a
diffusing particle. We denote the dynamical friction coefficient
(per unit mass) $\gamma=\Gamma/m$. If the force can be derived from
a potential, $F=-\nabla U(x)$, Langevin's equation\index{Langevin's
equation} takes the form
\begin{align}
m\ddot x+\Gamma \dot x+\nabla U(x)=\Xi.\label{Lang}
\end{align}

The main mathematical difference between the two approaches is that
Einstein assumes that the displacements $\Delta$ are independent,
whereas Langevin\index{Langevin's equation} assumes that the random
force\index{random force} $\Xi$ and the displacement $x$ are
independent. The two theories are reconciled in Section
\ref{s:displacement} below.

To investigate the statistical properties of the fluctuating force
$\Xi$, Langevin made the following assumptions.\newline

\noindent (i) \textit{The fluctuating force $\Xi$ is independent of the
velocity $v$.}\newline

\noindent (ii) \textit{$\Xi$ changes much faster than $v$.}\newline

\noindent (iii) $\langle \Xi\rangle=0$\newline

\noindent (iv) \textit{The accelerations imparted in disjoint time
intervals $%
\Delta t_1$ and $\Delta t_2$ are independent.}\newline

These conditions define the noise $\Xi(t)$ (so called \textit{white noise}) as the nonexistent derivative of Einstein's Brownian motion $X(t)$. The conditional probability distribution function of the velocity
process\index{velocity process} of a Brownian particle (PDF), given
that it started with velocity $v_0$ at time $t=0$, is defined as
$P(v,t\,|\,v_0)=\Pr \left\{v(t)<v\,|\,v_0\right\}$ and the
conditional probability density function is defined by
$$p\,(v,t\,|\,v_0)=\frac{\p P(v,t\,|\,v_0)}{\p v}.$$ In higher dimensions, we
denote the displacement vector $\x =(x_1,x_2,\dots,x_d)^T$, the
velocity vector $\dot{\x}= \mb{v}=(v_1,v_2,\dots,v_d)^T$, the random
force\index{random force} vector
$\mb{\Xi}=(\Xi_1,\Xi_2,\dots,\Xi_d)^T$, the PDF
$$P(\mb{v},t\,|\,\mb{v}_0)= \Pr
\left\{v_1(t)<v_1,v_2(t)<v_2,\dots,v_d(t)<v_d\,|\,\mb{v}(0)=\mb{v}_0\right\},$$
and the probability density function (pdf)
$$p\,(\mb{v},t\,|\,\mb{v}_0)=\frac{\p^nP(\mb{v},t\,|\,\mb{v}_0)}{\p v_1\p
v_2,\dots\p v_d}.$$

The conditioning implies that the initial condition for the pdf is
$p\,(\mb{v},t\,|\,\mb{v}_0)\to \delta(\mb{v}- \mb{v}_0)$ as $t\to0$.
According to the Waterston-Maxwell theory, when the system is in
thermal equilibrium\index{equilibrium}, the velocities of free
Brownian particles\index{free Brownian particle} have the
Maxwell--Boltzmann pdf; that is,
\begin{align}
\lim_{t\to\infty}p\,(\mb{v},t\,|\, \mb{v}_0)=\left( \frac{m}{2\pi
kT}\right)^{3/2} \exp\!\left\{-\frac{m|\mb{v}|^2}{ 2kT}\right\}.
\label{tc}
\end{align}

The solution of the Langevin equation\index{Langevin's equation}
(\ref{LE1}) for a free Brownian particle is given by
\begin{align}
\mb{v}(t)=\mb{v}_0e^{-\gamma t}+\frac{1}{m}
\int\limits_0^te^{-\gamma (t-s)}\mb{\Xi}(s)\,ds.  \label{sol}
\end{align}
 The integral \eqref{sol}
makes sense, if it is integrated once by parts. However, if the factor multiplying $\mb{\Xi}$ in
\eqref{sol} is non-differentiable as well, e.g., if it is $\x(t)$, integration by parts is
insufficient to make sense of the stochastic integral and a more sophisticated definition is needed.
Such a definition was given by It\^o in 1944 \cite{Ito} (see also \cite{Schuss2}).

To interpret the stochastic integral in the one-dimensional \eqref{sol}, we make a short
mathematical digression on the definition of integrals of the
type $\int_0^tg(s)\Xi(s)\,ds$, where $g(s)$ is a deterministic
integrable function. Such an integral is defined as the limit of
finite Riemann sums  of the form
\begin{align}
\int\limits_0^tg(s)\Xi(s)\,ds=\lim_{\Delta s_i\to0} \sum_ig(s_i)
\Xi(s_i)\,\Delta s_i,  \label{RS}
\end{align}
where $0=s_0<s_1<\cdots<s_N=t$ is a partition of the interval $[0,t]$.
According to the assumptions about $\Xi$, if we choose $\Delta s_i=\Delta
t=t/N$ for all $i$, the Gaussian increments $\Delta b_i=\Xi(s_i)\,\Delta s_i$ are independent identically
distributed (i.i.d.) random variables.
Einstein's  observation (see the beginning of Section
\ref{s:velocity}) that the RMS velocity on time intervals of length $\Delta t$
are inversely proportional to $\sqrt{\Delta t}$, implies that if the
increments $b_i=\Xi(s_i)\,\Delta s_i$ are chosen to be
normally distributed, their mean must be zero and their covariance
matrix must be $\langle\Delta b_i\Delta b_j\rangle=
q\Delta t\delta_{ij}$ with $q$ a parameter to be determined. We write $\Delta
b_i \sim{\cal N}\left(0,{q\Delta t}\right)$. Then $g(s_i)\Xi(s_i)\,\Delta
s_i\sim {\cal N}\left(0,|g(s_i)|^2{q\Delta t}\right)$, so that
$\sum_ig(s_i)\Xi(s_i)\,\Delta s_i \sim{\cal N}(0,\sigma_N^2)$, where
$\sigma_N^2=\sum_i|g(s_i)|^2q\Delta s_i$. As $\Delta t\to 0$, we obtain
$\lim_{\Delta t\to 0}\sigma_N^2=q\int_0^tg^2(s)\,ds$ and $\ds
\int_0^tg(s)\Xi(s)\,ds\sim{\cal N} (0,\sigma^2)$, where
\begin{align}
\sigma^2=q\int\limits_0^tg^2(s)\,ds.  \label{s2}
\end{align}

To interpret \eqref{sol}, we use \eqref{s2} with $g(s)=e^{-\gamma(t-s)}$ and obtain
\begin{align}
\sigma^2=\frac{q}{2\gamma}\left(1-e^{-2\gamma t}\right).
\label{sig2}
\end{align}
Returning to the velocity vector $\mb{v}(t)$, we obtain from the
above considerations
\begin{align}
\mb{v}(t)-\mb{v}_0e^{-\gamma t}\sim{\cal N}
\left(\mb{0},\sigma^2\mb{I}\right)  \label{v-v0}
\end{align}
with $\sigma^2$ given by \eqref{sig2}. Finally, the condition
(\ref{tc}) implies that $q=2\gamma kT/m$, so that in 3-D the mean
energy is as given in eq. (\ref{WM}). It\^o's construction of the stochastic integral allows $g(t)$ to be stochastic, but independent of the Brownian increments $\Xi(s_i)\Delta s_i=X(s_i+\Delta s_i)-X(s_i)$.

In the limit $\gamma\to\infty$ the acceleration $\gamma\mb{v}(t)$
inherits the properties of the random acceleration $\mb{\Xi}(t)$ in
the sense that for different times $t_2>t_1>0$ the accelerations
$\gamma\mb{v}(t_1)$ and $\gamma\mb{v}(t_2)$ become independent. In
fact, from eqs.(\ref{sol}) and  $\Delta
b_i \sim{\cal N}\left(0,{q\Delta t}\right)$, we find that
\begin{align}
\lim_{\gamma\to\infty}\eE[\gamma\mb{v}(t_1)\cdot \gamma\mb{v}(t_2)]\label{vdelta}=0\nonumber
\end{align}
and a similar result for $0<t_2<t_1$. It follows that for $\gamma (t_1\wedge t_2)\gg1$,
 \begin{equation}
\eE[\gamma\mb{v}(t_1)\cdot \gamma\mb{v}(t_2)]=\frac{\gamma q}{m^2}e^{-\gamma|t_2-t_1|}(1+o(1))
=\frac{2q}{m^2}\delta(t_2-t_1)(1+o(1)),\label{deltav}
\end{equation}
because for $t_1>0$
\begin{align}
\lim_{\gamma\to\infty}\int\limits_0^\infty f(t_2)\frac{\gamma}{m^2}e^{-\gamma|t_2-t_1|}(1+o(1))\,dt_2=\frac{2}{m^2}f(t_1)
 \end{align}
 for all test functions $f(t)$ in ${\rR}^{+}$.

\subsection{The displacement process\index{displacement process}\label{s:displacement}}

The displacement of a free Brownian particle\index{free Brownian
particle} is obtained from integration of the velocity
process\index{velocity process},
\begin{align}
\x(t)=\x_0+\int\limits_0^t\mb{v}(s)\,ds. \label{xv}
\end{align}
Using the expression (\ref{sol}) in eq. (\ref{xv}) and changing the
order of integration in the resulting iterated integral, we obtain
\begin{align}
\x(t)-\x_0-\mb{v}_0\frac{1-e^{-\gamma
t}}{\gamma}=\int\limits_0^tg(s) \mb{\Xi} (s)\,ds,  \label{xt}
\end{align}
where $g(s)=(1-e^{-\gamma(t-s)})/m\gamma$.

Reasoning as above, we find that the stochastic integral in eq.
(\ref{xt}) is a normal variable with zero mean and covariance
matrix\index{covariance matrix} $\mb{\Sigma}=\sigma^2\mb{I}$, where
\begin{align}
\sigma^2=q\int\limits_0^tg^2(s)\,ds=\frac{q}{2\gamma^3}\left(2\gamma
t-3+4e^{-\gamma t}-e^{-2\gamma t}\right).  \label{S2}
\end{align}
The moments of the displacement are
\begin{align}
\eE\left[ \x (t)-\x _0-\mb{v}_0\frac{1-e^{-\gamma t}} {\gamma}
\right]=0 \label{md}
\end{align}
and the conditional second moment of the displacement is
\begin{eqnarray}
&&\,\eE\left(\left|\x (t)-\x_0\right|^2\,|\,\x
_0,\mb{v}_0\right)=\int|\x -
\x _0|^2p\,(\x,t\,|\, \x_0,\mb{v}_0)\,d\x   \nonumber  \\
%&\nonumber\\
&=&\,\frac{|\mb{v}_0|^2}{\gamma^2}\left(1-e^{-\gamma t}\right)^2+
\frac{3kT}{m\gamma^2}\left(2\gamma t-3+4e^{-\gamma t}-e^{-2\gamma
t}\right),\label{cvariance}
\end{eqnarray}
which is independent of $\x_0$. Using the Maxwell distribution of
velocities (\ref{tc}), we find that the unconditional second moment
is
\begin{align}
\eE\left|\x(t)-\x _0\right|^2 =&\,\eE_{\x_0}\eE_{\mb{v}_0}
\left(\left|\x(t)-\x_0\right|^2\,|\,\x_0,\mb{v}_0\right)  \nonumber \\
%&\nonumber\\
=&\,\frac{3kT}{m\gamma^2} \left(1-e^{-\gamma
t}\right)^2+\frac{3kT}{m\gamma^2}\left(2\gamma t-3+4e^{-\gamma t}-
e^{-2\gamma t}\right)\nonumber\\
%&\nonumber\\
=&\,\frac{6kT}{m\gamma^2} \left(\gamma t-1+e^{-\gamma t}\right).
\label{varx}
\end{align}

The long time asymptotics of $\eE\left|\x(t)-\x _0\right|^2$ is
found from \eqref{varx} to be
\begin{align}
\eE\left|\x (t)-\x _0\right|^2
\sim\frac{6kT}{m\gamma}t=\frac{kT}{ma\eta}t\hspace{0.5em}\mbox{for}\
t\gamma\gg1; \label{lta}
\end{align}
that is, the displacement variance of each component is
asymptotically $kT/3ma\eta$. It was this fact that was verified
experimentally by Perrin \cite{Perrin}\index{Perrin, J.}. The
one-dimensional diffusion coefficient\index{diffusion coefficient},
as defined in \eqref{moments}, is therefore given by
$D=kT/6ma\eta$.

Equation (\ref{varx}) implies that the short time asymptotics of $\eE\left|%
\x (t)-\x _0\right|^2$ is given by
\begin{align}
\eE\left|\x (t)-\x _0\right|^2
\sim\frac{3kT%
}{m}t^2=\langle|\mb{v}_0|^2\rangle t^2.  \label{sta}
\end{align}
This result was first obtained by Smoluchowski.

\subsection{Reconciliation of Einstein's and Langevin's theories}To reconcile the Einstein\index{Einstein, A.} and the
Langevin approaches, we have to show that
for two disjoint time intervals, $(t_1,t_2)$ and $(t_3,t_4)$, in the
limit $\gamma\to\infty$, the increments
$\Delta_1\x = \x (t_2)- \x(t_1)$ and $\Delta_3 \x =\x (t_4)-\x(t_3)$
are independent zero mean Gaussian variables with variances proportional to the time
increments. Equation (\ref{xt}) implies that
in the limit $\gamma\to\infty$ the increments $\Delta_1\x $ and $\Delta_3\x $ are zero mean Gaussian
variables and \eqref{lta} shows that
the variance of an increment\index{Random increment} is proportional
to the time increment\index{Random increment}.

To show that the increments are independent, we use \eqref{deltav} in \eqref{xt} to obtain
\begin{align}
\lim_{\gamma\to\infty}
\gamma^2\langle\Delta_1\x\cdot\Delta_3\x\rangle=&\,
\lim_{\gamma\to\infty}\int\limits_{t_1}^{t_2}\int\limits_{t_3}^{t_4}
\langle\gamma \mb{v}(s_1) \cdot\gamma\mb{v}(s_2)\rangle\,ds_1\,ds_2
\nonumber \\
%&\nonumber\\
=&\,\frac{2q}{m^2}\int\limits_{t_1}^{t_2}\int\limits_{t_3}^{t_4}\delta(s_2-s_1)
\,ds_1\,ds_2=0.  \label{dxindep}
\end{align}
As is well-known \cite{Feller}, uncorrelated
Gaussian variables\index{Gaussian variable} are independent. This
reconciles the Einstein\index{Einstein, A.} and Langevin
theories\index{Langevin's equation} of Brownian
motion\index{Brownian motion} in liquid.

Introducing the dimensionless variables $s=\gamma t$ and $\mb{\xi}(s)=\sqrt{m/6kT} \gamma\x(t)$, we find from
\eqref{varx} that
\begin{align}
\lim_{\gamma\to\infty}\eE\left|\mb{\xi}(s)- \mb{\xi}
(0)\right|^2=s-1+e^{-s}\sim s\hspace{0.5em}\mbox{for}\  s\gg1
\label{varxi}
\end{align}
and from \eqref{dxindep} that
\begin{align}
\lim_{\gamma\to\infty}\eE[\Delta_1\mb{\xi}\cdot \Delta_3\mb{\xi}]=0.
\label{indepxi}
\end{align}
Equations (\ref{varxi}) and (\ref{indepxi}) explain (in the context
of Langevin's description) Einstein's quoted assumption that ``\textit{... the
movement of one and the same particle after different intervals of
time [are] mutually independent processes, so long as we think of
these intervals of time as being chosen not too small.}''

\section{Smoluchowski's  limit of Langevin's equation}\label{s:Smol}
The Langevin equation \eqref{Lang} serves as a model
for many activated processes \cite{Landauer}, for which the
escape rate determines their time evolution. It is one of the most extensively
studied equations in statistical physics \cite{HTB}. For high damping, the joint probability density
function (pdf) of displacement $x(t)$ and velocity $\dot x(t)$ breaks into a product of
the stationary Maxwellian pdf of the velocity $v=\dot x$ and the time-dependent
pdf of the displacement $x(t)$, which satisfies an altogether different equation. This is the case
not only for the linear Langevin equation, but holds in general.

%\subsection{The Smoluchowski limit}\label{t:Smoluchowski}
Smoluchowski has shown \cite{Smoluchowski} that as $\gamma\to\infty$ the trajectories $x(t)$ of
the Langevin equation \eqref{Lang} converge in probability to these of the Smoluchowski equation
\begin{equation}
\gamma \dot{x}+U^{\prime }(x)=\sqrt{2\eps\gamma}\,\dot{w},
\label{LSE12}
\end{equation}
where $\dot w(t)=\Xi(t)$ is $\delta$-correlated Gaussian white noise.

A more modern derivation begins with writing the Langevin equation \eqref{Lang} as the phase space
system
\begin{align}
\dot{x} =&\,v  \label{xd} \\
%&\nonumber\\
 \dot{v} =&\,-\gamma v-U^{\prime
}(x)+\sqrt{2\eps\gamma}\,\dot{w}, \label{yd}
\end{align}
and with scaling time by setting
\begin{equation}
t=\gamma s  \label{tgs}.
\end{equation}
The scaled Brownian motion
$w(t)=\sqrt{\gamma} w^\gamma (s)$, $w^\gamma (s)$ is a standard Brownian
motion in time $s$. The scaled white noise is formally
$$\dot{w}(t)=\gamma^{-1}d\sqrt{\gamma }w^\gamma
(s)/ds=\gamma^{-1/2}\stackrel{\circ }{w^\gamma}(s).$$

Setting $x^\gamma (s)=x(\gamma t)$, $v^\gamma (s)=v(\gamma t)$, and using the overcircle notation for the derivative with respect to $s$, we note
that $\dot{x}(t)=\gamma^{-1} \stackrel{\circ }{x^\gamma}(s)$,
$\dot{v}(t)=\gamma^{-1} \stackrel{\circ }{v^\gamma} (s)$,
\eqref{yd} takes the form $$\stackrel{\circ}{v^\gamma}
(s)+\gamma^2v^\gamma (s) =-\gamma U'(x^\gamma (s))+\gamma
\sqrt{2\eps}\stackrel {\circ }{w^\gamma}(s).$$ Hence
\[ v^\gamma (s)=v^\gamma (0)e^{-\gamma ^2s}+\gamma \int\limits_0^se^{-\gamma
^2(s-u)}\left[ -U^{\prime }(x^\gamma
(u))\,du+\sqrt{2\eps}\,dw^\gamma (u)\right], \] which can be written as
 $\dot{x}(t)=\gamma^{-1} \stackrel{\circ
}{x^\gamma}(s)=v^\gamma (s)$, so that
\begin{align*}
x^\gamma (s) =&\,x^\gamma (0)+\gamma \int\limits_0^sv^\gamma
(z)\,dz=x^\gamma
(0)+v^\gamma (0)\frac{1-e^{-\gamma ^2s}}\gamma  \\
%&\nonumber\\
&\,+\gamma ^2\int\limits_0^s\int\limits_0^ze^{-\gamma ^2(z-u)}\left[
-U'(x^\gamma (u))\,du+\sqrt{2\eps}\,dw^\gamma (u)\right] \,dz.
\end{align*}
In the limit $\gamma\to\infty$
\[x^\infty (s)= x^\infty (0)+\int\limits_0^s\left[ -U^{\prime
}(x^\infty (u))\,du+\sqrt{2\eps}\,dw^\infty (u)\right],
\]
where $w^\infty(u)$ is Brownian motion. This is the integral form of the Smoluchowski stochastic differential equation
\begin{equation}
\stackrel{\circ}{x^\gamma}(s)=-U^{\prime }(x^\infty (s))+\sqrt{2\eps}\,\stackrel{\circ}{w^\infty}
(s).  \label{LSE}
\end{equation}
Returning to the original time scale, \eqref{LSE} becomes
\eqref{LSE12}, which means that \eqref{LSE} is the Langevin
equation \eqref{Lang} without the inertia term $\ddot{x}$. This means that the limit exists on every
trajectory.

\subsection{Numerical solution of the Smoluchowski equation}\label{ss:numerical}

Starting with the Smoluchowski equation in dimension $d\geq1$
\begin{align}
\dot \x(t)=\mb{a}(\x(t),t)+\mb{b}(\x(t),t)\,\dot\w(t),\label{Seq}
\end{align}
where $\x$ and $\mb{a}(\x,t)$ are $d$-dimensional vectors, $\w(t)$ is an $m$-dimensional Brownian
motion, and $\mb{b}(\x,t)$ is a $d\times m$ matrix, approximate trajectories of \eqref{Seq} can be constructed on any time interval $s<t<T$ by
the Euler scheme
 \begin{align}
\x(t+\Delta t)=&\,\x(t)+\mb{a}(\x(t),t)\Delta
t+\mb{b}(\x(t),t)\,\Delta \w(t,\Delta
t)\label{EulerRd2}\\
\x_N(s)=&\,\x_0.\nonumber
 \end{align}

Boundary behavior can be imposed on the trajectories of \eqref{EulerRd2} in a given domain $D$. For
example, all trajectories of \eqref{EulerRd2} can be instantaneously
terminated when they exit $D$. In this case the boundary $\p D$ is called \textit{an absorbing
boundary.} The trajectories can be instantaneously reflected at $\p D$ back into $D$ according to a
given reflection law; in this case $\p D$ is called \textit{a reflecting boundary.} They can also
be either instantaneously terminated with a given probability or instantaneously reflected; then $\p
D$ is called \textit{a partially reflecting boundary.}

It can be shown that the trajectories of
\eqref{EulerRd2} with a given boundary behavior converge to a limit  as $\Delta t\to0$. The limit
is defined as the solution of the Smoluchowski equation \eqref{Seq} with the given boundary
behavior. (see \cite{Schuss2} and \cite{Schuss4} for details and for other types of boundary behavior.)

\subsection{The probability density of the Smoluchowski trajectories }

%\section[The Probability Density of Euler's Scheme in $\rR$ and the FPE]{The pdf\index{Probability density function} of Euler's Scheme in $\rR$ and the FPE\index{Fokker--Planck equation}}\label{s:FPEPI}
%%%%%%%%%%%%%%%%%%%%%%%%%%%%%%%%%%%%%%%%%%%%%%%%%%%%%%%%%%
\font\bb=msbm10 at 11pt \font\bbbis=msbm10 at 10pt \font\tenbb=msbm10
 \def\rR{\hbox{\bb R}}
We assume that in the one-dimensional case $b(x,t)>\delta>0$  for some constant $\delta$. We assume for now that $a(x,t)$ and $b(x,t)$ are deterministic functions. To construct the pdf of a trajectory of \eqref{EulerRd2}, we note that the pdf of $x_N(t)$ can be expressed
explicitly for $t$ on the lattice, because \eqref{EulerRd2}, written as
\begin{align}
\Delta w(t)=\frac{x_N(t+\Delta
t)-x_N(t)-a(x_N(t),t)}{b(x_N(t),t)},\label{Deltaw}
\end{align}
means that for all $t$ on the lattice the expressions on the right-hand side of
\eqref{Deltaw} are independent, identically distributed Gaussian variables. It follows  that the pdf of the entire Euler trajectory is the product
\begin{align}
&\,p{\Big(}x_1,t_1;x_2,t_2;\dots;x_n,t_n{\Big)}\label{Epdf}\\
&=\,\prod_{k=1}^n\left[2\pi b^2(x_{k-1},t_{k-1})\Delta t
\right]^{-1/2}\!\exp\left\{-\frac{[x_k-x_{k-1}-a(x_{k-1},t_{k-1})\Delta
t]^2}{2b^2(x_{k-1},t_{k-1})\Delta t} \right\}.\nonumber
\end{align}
Setting $x_n=x$ and integrating over $\rR$ with respect to all intermediate
points $x_1,x_2,\,\ldots\,,x_{n-1}$, we find from \eqref{Epdf} that the
transition pdf of the trajectory satisfies on the lattice the recurrence relation
\begin{align}
p_{N}(x,t+\Delta t\,|\,x_{0})=\int\limits_{\rR} \frac{
p_{N}(y,t\,|\,x_{0})\,dy}{\sqrt{2\pi \Delta t}\,b (y,t)}\!\exp \left\{
-\frac{\left[ x-y-a(y,t)\Delta t\right]^{2}}{ 2b ^{2}(y,t)\Delta t}\right\}
.\label{PN+1t}
\end{align}
The solution of the integral equation \eqref{PN+1t} is called Wiener's \textit{discrete path integral}. Its limit as $N\to\infty$ is called Wiener's \textit{path integral}.

For $\Delta t=(t-s)/N$, in the limit $N\to\infty$, the pdf $p_{\scriptsize N}(x,t\,|\,x_{0})$ of the solution $x_N(t,{\mathfrak{w}})$ of \eqref{EulerRd2}
converges to the solution $p\,(x,t\,|\,x_0)$ of \eqref{PN+1t}, where ${\mathfrak{w}}$ is the entire discrete path of the Brownian motion $w(t)$ on the lattice.  Expansion of $p_{\scriptsize N}(x,t\,|\,x_{0})$ in \eqref{PN+1t} shows that the pdf $p(x,t\,|\,x_{0})$ is also the solution of the initial value problem
\begin{equation}
\frac {\p  p\,(y,t\,|\,x,s)} {\p  t}= \frac 12\frac{\p ^2\left[
b^2(y,t)p\,(y,t\,|\,x,s)\right]}{\p  y^2} -\frac{\p  \left[
a(y,t)p\,(y,t\,|\,x,s)\right] }{\p  y} \label{1DFPE}
\end{equation}
with the initial condition
\begin{equation}
\lim_{t\downarrow s}p\,(y,t\,|\,x,s)=\delta(y-x).\label{1dic}
\end{equation}
Equation \eqref{1DFPE} is called the \textit{Fokker-Planck-(Smoluchowski) equation (FPE)} (see \cite{Schuss4}).

In the Smoluchowski equation \eqref{Seq} for $d=n\geq1$, the $n\times m$ matrix
\begin{align}
\mb{B}\left(\x,t\right)=\left\{b^{ij} \left(\x,t\right)\right\}
_{n\times m}\label{sbbT}
\end{align}
is called the \textit{noise matrix},  and
$\mb{\sigma}\left(\x,t\right) =\frac 12\mb{B}\left(\x,t\right)\mb{B}^T\left(\x,t\right)$
is called the \textit{diffusion matrix}. The operator
\begin{align}
\ds{{\cal L}_{\small{\y}}}p=\sum_{i=1}^d\frac\p {\p  y^i}\left\{
\sum_{j=1}^d\frac{\p }{\p  y^j}\sigma^{ij} (\y,t)
p-a^i\left(\y,t\right) p \right\},\label{FPO}
\end{align}
 is called the \textit{Fokker--Planck operator}, or the \textit{forward Kolmogorov operator.}

As in the one-dimensional case, the pdf
$p\left(\y,t\,|\,\x,s\right)$ that $\x(t)=\y$, given that $\x(s)=\x$, is the solution of the Fokker-Planck-Smoluchowski initial value problem
\begin{align}
\frac{ \p  p\left(\y,t\,|\,\x,s\right)}{\p  t} =&\, \ds{{\cal
L}_{\small{\y}}}p\left(\y,t\,|\,\x,s\right)\hspace{0.5em}\mbox{ for}\
\x,\y\in\rR^n,\ t>s,\label{FPE}\\
\lim_{t\downarrow s} p\left(\y,t\,|\,\x,s\right)=&\,\delta(\x-\y).
\label{strongIC}
\end{align}
It can be written as the conservation law by defining the probability flux density vector
 \begin{align}
 J^i(\y,t\,|\,\x,s)=-\sum_{j=1}^d\frac{\p }{\p  y^j}\sigma^{ij} (\y,t)
p\left(\y,t\,|\,\x,s\right)-a^i\left(\y,t\right) p\left(\y,t\,|\,\x,s\right)
\end{align}
and writing \eqref{FPE} in the divergence form
\begin{align}
\frac{ \p  p\left(\y,t\,|\,\x,s\right)}{\p  t}=-\mbox{div}_{\y}\,\mb{J}(\y,t\,|\x,s).\label{div}
\end{align}

The probability density at time $t$ and at point $\x$ can be represented for any domain $\Omega$ by the limit as $N\to\infty$ of
\beq
&&\Pr\Big{\{}{\x}_N(t_{1,N})\in\Omega,{\x}_N(t_{2,N})\in\Omega,\dots,
{\x}_N(t)=\x, t\leq T\leq t+\Delta t\,|\,\x(0)=\y\Big{\}}\nonumber=\\
&&\nonumber\\
&&\Bigg{[}\int_{\Omega} \int_{\Omega}\cdots
\int_{\Omega}\,\prod_{j=1}^{N} \frac{d{\y}_j}{\sqrt{(2\pi \Delta
t)^n\det\mb{\sigma}(\x)(t_{j-1,N}))}}
\nonumber\\
&&\nonumber\\
 &&\times \exp \Bigg{\{} -\frac{1}{2\Delta t}
\left[\mb{\y}_j-\x(t_{j-1,N})- \mb{a}({\x}(t_{j-1,N}))\Delta t
\right]^T\mb{\sigma}^{-1}(\x(t_{j-1,N}))
\nonumber\\
&&\nonumber\\
&&\times\left[{\y}_j-\x(t_{j-1,N})-\mb{a}(\x(t_{j-1,N}))\Delta t
\right]\Bigg{\}} ,\label{Wiener1}
 \eeq
 where
\begin{eqnarray*}
\Delta t=\frac{t}{N},\quad t_{j,N}=j\Delta t,
\end{eqnarray*}
and
\[\x(t_{0,N})=\y\]
in the product. The limit is the Wiener integral defined by the
stochastic differential equation (\ref{Seq}) with appropriate boundary condition. In the limit $N\to\infty$ the integral
(\ref{Wiener1}) converges to the solution of the Fokker-Planck
equation (\ref{div}) in $\Omega$. Similarly, expanding the path integral with respect to $\x$ and $s$, we find that $p\left(\y,t\,|\,\x,s\right)$ satisfies the \textit{backward Kolmogorov equation} \cite{Schuss2}
\begin{align}
{\cal L}^*_{\footnotesize{\x}} p\left(\y,t\,|\,\x,s\right) =\sum_{i=1}^n\sum_{j=1}^n
\sigma^{ij}\left(\x,s\right) \frac{\p ^2p\left(\y,t\,|\,\x,s\right)} {\p x^i\p
x^j}+ \sum_{i=1}^na^i\left(\x,s\right) \frac{\p p\left(\y,t\,|\,\x,s\right)}{\p
x^i} \label{BKO}
\end{align}
with the terminal condition
\begin{equation}
\lim_{t\uparrow s}p\,(\y,t\,|\,\x,s)=\delta(\y-\x).\label{1dic2}
\end{equation}
The Fokker-Planck equation can be generalized when a killing measure is added to the dynamics. A killing measure represents the probability per unit time and unit length to terminate a trajectory at a given point at a given time. Mixed boundary conditions and Fokker-Planck equation for the survival probability with killing is discussed in \cite{HolcmanSchuss2015}.

%%%%%%%%%%%%%%%%%%%%%%%%%%%%%%%%%%%%%%%%%%
\subsection{Diffusion processes and diffusion models of large data sets}\label{diffusion}
%%%%%%%%%%%%%%%%%%%%%%%%%%%%%%%%%%%%%%%%%%
A $d$-dimensional Markov process $\x (t)$ is called a \textit{diffusion process} with (deterministic) drift vector field $\mb{a}(\x,t)$ and \textit{(deterministic) diffusion matrix}
$\mb{\sigma}(\x,t)$, if it has continuous trajectories,
\begin{align}
&\lim_{\Delta t\rightarrow 0}\frac 1{\Delta t}\eE\left\{ \x
(t+\Delta
t)-\x (t)\,|\,\x (t)= \x \right\} =\mb{a}(\x,t)\label{Drift}\\
%&\mbox{}
&\lim_{\Delta t\rightarrow 0}\frac 1{\Delta t}\eE\left\{ \left[
x^i(t+\Delta t)-x^i(t)\right]\left[ x^j(t+\Delta
t)-x^j(t)\right]\,|\,\x (t)=\x \right\}\nonumber\\
=&\,{{\sigma}}^{ij}(\x,t) \label{sijt}
\end{align}
for $i,j=1,2,\ldots,d$, and for some $\delta >0$
\[
\lim_{\Delta t\rightarrow 0}\frac 1{\Delta t}\eE\left\{ \left| \x
(t+\Delta t)-\x (t)\right| \,^{2+\delta }\,|\,\,\x (t)=\x \right\}
=0.
\]
It can be shown \cite{Schuss2} that under mild regularity assumptions on the coefficient the solution of the stochastic differential equation (SDE)
\begin{align}\label{stochlocal01}
\dot{\x}(t)=\mb{a}(\x(t),t) +\mb{B}(\x(t),t)\, \dot{\w}(t)
\end{align}
is a diffusion process with drift field $\mb{a}(\x,t)$ and diffusion matrix
$\mb{\sigma}(\x,t)=\frac12\mb{B}(\x,t)\mb{B}^T(\x,t)$. Also a partial converse is true: given sufficiently nice drift field $\mb{a}(\x,t)$ and a strictly positive definite diffusion matrix $\mb{\sigma}(\x,t)$ of a diffusion process, there is a noise matrix $\mb{B}(\x,t)$ and a Brownian motion $\w(t)$ such the process is a solution of the SDE \eqref{stochlocal01} \cite{Karlin,Schuss2}.

Equations \eqref{Drift} and \eqref{sijt} reconstruct the SDE from its trajectories. The discrete version of these equations can be used as approximations to the coefficients, when a sufficiently large set of trajectory fragments is given (see \cite{HozeBJ2014,holcmanBJ2015,holcmanhoze2015}). Indeed, large number (tens of thousands) of short single particle trajectories (SPTs), collected by super-resolution methods, at tens of nanometer precision for motion occurring in cell are ideal data to recover drift and diffusion tensors.  But the sampled molecular process cannot be directly modeled by the overdamped Langevin equation, which describes diffusion on the microscopic level, because an additional noise of localization is added in tracking the motion (algorithm and point spread function error).  Using a stochastic model of the acquired data to calibrate the model, it is possible to distinguish the perturbation added to the physical motion (stochastic equation) \cite{holcmanhoze2015}. The denoising procedure of the SPTs reveals that effective diffusion coefficient contains the divergence of deterministic drift component and also provides a criteria to differentiate trapped stochatic particle from immobile ones \cite{holcmanhoze2015}.

\subsection{The MFPT and survival probability}
%%%%%%%%%%%%%%%%%%%%%%%%%%%%%%%%%%%%%%%%%%
If the Smoluchowski trajectories are terminated at the boundary $\p D$ of a given domain, the pdf vanishes on $\p D$. Thus equations \eqref{FPE} and \eqref{strongIC} have to be supplemented by the boundary condition
\begin{align}
p\left(\y,t\,|\,\x,s\right)=0\hspace{0.5em}\mbox{ for}\ \y\in \p D,\ \x \in D,\ t>s.
\end{align}
In this case, the first passage time to the boundary $\p D$ is
\begin{align}
\tau_{D}=\inf\{t:\, \x(t)\not\in D\}.
\end{align}
Thus, the probability that the Smoluchowski trajectory is still in $D$ at time $t>s$, given that {at time $s<t$} it started at $\x\in D$, is given by
\begin{align}
\Pr\{\x(t)\in D\,|\,\x(s)=\x\}=\int\limits_Dp(\y,t\,|\,{\x,s})\,d\y\label{survival}
\end{align}
Actually, $\Pr\{\x(t)\in D\,|\,\x(s)=\x\}=\Pr\{\tau_D>t\,|\,\x(s)=\x\},$  which is \textit{the survival probability} at time $t>s$
of the Smoluchowski trajectory that started at $\x$ {at time $s<t$.}

If the Smoluchowski trajectories are reflected at the boundary $\p D$ of a given domain such that the normal flux at the boundary
vanishes \cite{Schuss1}, then the boundary condition becomes
\begin{align}
\mb{J}(\y,t\,|\,\x,s)\cdot\mb{\nu}(\y)=0\hspace{0.5em} \mbox{for} \y \in\p D,\x\in D,\ t>s,
\end{align}
where $\mb{\nu}(\y)$ is the unit outer normal vector  at $\y\in\p D$.

When the boundary $\p D$ absorbs the Smoluchowski trajectories, the expected MFPT to the boundary of Smoluchowski trajectories in
$D$ is found by integrating the survival probability to obtain the MFPT as
\begin{align}
\eE[\tau_D\,|\,\x(s)=\x]=\int\limits_s^\infty\int\limits_Dp(\y,t\,|\,{\x,s})\,d\y\,dt.\label{MFPT}
\end{align}
Setting $u(\x,s)=\eE[\tau_D\,|\,\x(s)=\x]$, using the backward Kolmogorov equation and Green's identity, we obtain the backward
boundary value problem
\begin{align}
{{\cal L}^*}_{\x}u(\x,s)+\frac{\p u(\x,s)}{\p s}=&-1\hspace{0.5em}\mbox{for}\ \x\in D,\label{PAVs}\\
u(\x,s)=&\,0\hspace{0.5em}\mbox{for}\ \x\in \p D.\label{Kbc}
\end{align}
Note that if the coefficients $\mb{\sigma}$ and $\mb{a}$ are time-independent, then \eqref{BKO}, \eqref{Kbc} reduce to the time-homogeneous elliptic Pontryagin-Andronov-Vitt (PAV) \cite{PAV} boundary value problem in $D$
\begin{align}
{{\cal L}^*}_{\x}u(\x)=&-1\hspace{0.5em}\mbox{for}\ \x\in D,\label{PAV}\\
u(\x)=&\,0\hspace{0.5em}\mbox{for}\ \x\in \p D.\label{hbc}
\end{align}
If the boundary is reflecting, then the absorbing boundary condition \eqref{hbc} for the PAV equation \eqref{PAV} is changed to
\begin{align}
\mb{J}(\x)\cdot\mb{\nu}(\x)=\,0\hspace{0.5em}\mbox{for}\ \x\in \p D.\label{rbc}
\end{align}

If $\p D$ is absorbing on a part $\p D_a$ and reflecting on the remaining part $\p D_r-\p D-\p D_a$, then the boundary conditions for
the PAV equation is absorbing on $\p D_a$ and reflecting on $\p D_b$.

Note further that the MFPT to the boundary is given by \eqref{MFPT} , where $p(\y,t\,|\,{\x,s})$ is the solution of the FPE with
absorbing boundary conditions also in the case of reflecting boundaries, because prior to reaching the boundary the Smoluchowski
trajectories are independent of boundary behavior.

%%%%%%%%%%%%%%%%%%%%%%%%%%%%%%%%%%%%%%%%%%
\section{Modeling in cell using the stochastic Narrow Escape}
%%%%%%%%%%%%%%%%%%%%%%%%%%%%%%%%%%%%%%%%%%
The narrow escape problem is to evaluate the MFPT when the reflecting part of the boundary $\p D_b$ is much bigger than the absorbing part $\p D_a$ \cite{Ward1,Ward2,HS1,PNAS1}, \cite{NarrowEscape1,NarrowEscape2,NarrowEscape3,Coombs,Cheviakov}. In this case $\p D_a$ represents a small absorbing window in the boundary, through which trajectories can escape the domain $D$, while the large reflecting part $\p D_b$ represents an impermeable wall (Fig. \ref{f:trajectory}), such as a lipid cell membrane that is impermeable to diffusing ions. This mathematical model represents many biological models. Thus equation \eqref{PAV} and the boundary conditions \eqref{hbc} or \eqref{rbc} are the  basis for the analysis of the narrow escape problem.

The MFPT $\bar\tau$ depends on the starting point $\x$ of the Brownian trajectory, thus it should
be denoted $\bar\tau(\x)$. This function is the solution of the classical mixed
Neumann-Dirichlet boundary value problem for the Laplace equation \cite{Schuss2,HolcmanSchuss2015},
\beq
D\nabla^2\bar\tau(\x)&=&-1\hspace{0.5em}\mbox{for}\ \x\in\Omega\label{eq}\\
\frac{\p\bar\tau(\x)}{\p n}&=&0\hspace{0.5em}\mbox{for}\ \x\in\p\Omega_r\label{Neumann}\\
\bar\tau(\x)&=&0\hspace{0.5em}\mbox{for}\ \x\in\p\Omega_a,\label{Dirichlet}
\eeq
where $D$ is the diffusion coefficient and $\n$ is the unit outer normal to the boundary
\cite{Schuss2}. {The system (\ref{eq})-(\ref{Dirichlet}) follows from the backward Kolmogorov equation
\cite{Schuss2} (the adjoint of the Fokker-Planck equation) for the transition probability density function
$p(\y,t\,|\,\x)$ of the Brownian trajectories,
 \beq
\frac{\p p(\y,t\,|\,\x)}{\p t}&=&D\nabla^2_{\x} p(\y,t\,|\,\x)\hspace{0.5em}\mbox{for}\ \x,\y\in
\Omega\\
\frac{\p p(\y,t\,|\,\x)}{\p n_{\x}}&=&0\hspace{0.5em}\mbox{for}\ \y\in \Omega, \x\in \Omega_r\label{N}\\
p(\y,t\,|\,\x)&=&0\hspace{0.5em}\mbox{for}\ \y\in \Omega, \x\in \Omega_a\label{D}\\
p(\y,0\,|\,\x)&=&\delta(\y-\x)\hspace{0.5em}\mbox{for}\ \x, \y\in \Omega.\label{I}
 \eeq}
{The survival probability of Brownian trajectories that start at $\x\in\Omega$ is
 \beq
\Pr\{\tau>t\,|\,\x\}=\int\limits_{\Omega}p(\y,t\,|\,\x)\,d\y
 \eeq
and its mean value is
 \beq
\bar\tau(\x)=\int\limits_0^\infty\Pr\{\tau>t\,|\,\x\}\,dt.
 \eeq
It follows that
 \beq
D\nabla^2\bar\tau(\x)=\int\limits_0^\infty D\nabla^2_{\x}\Pr\{\tau>t\,|,\x\}\,dt=
\int\limits_0^\infty\int\limits_{\Omega} \frac{\p p(\y,t\,|\,\x)}{\p t}\,d\y\,dt=-1.\label{DN}
 \eeq
The last equality in (\ref{DN}) follows from the initial condition (\ref{I}) and the Neumann and Dirichlet
conditions (\ref{Neumann}), (\ref{Dirichlet}) are inherited from  (\ref{N}) and (\ref{D}), respectively.}

No explicit solutions to the problem (\ref{eq})--(\ref{Dirichlet}) are known {in general} \cite{HolcmanSchuss2015}. If the
absorbing {part of the} boundary $\p\Omega_a$ is much smaller than the entire boundary $\p\Omega$, numerical solutions to the problem are
very hard to construct due to the presence of a boundary layer near $\p\Omega_a$, where gradients are very large so the numerical complexity
becomes prohibitive. The problem cannot be circumvented by Brownian dynamics simulations of the MFPT $\bar\tau$, because reaching $\p\Omega_a$ is a rare
event on the time scale of diffusion. The remedy to these difficulties is the construction of analytical approximations to the solution of
(\ref{eq})--(\ref{Dirichlet}) by new asymptotic methods developed specifically for the problem at hand. In the next section, we summarize the asymptotic formulas solution of (\ref{eq})--(\ref{Dirichlet}). We briefly mention how they are derived and refer to \cite{HolcmanSchuss2015} for more details.

%%%%%%%%%%%%%%%%%%%%%%%%%%%%%%%%%%%%%%%%%%%%%%%%%%%%%%%%%%%%%%%%%%%%%%%%%%%%%%%%
\subsection{Narrow escape formula in two-dimensions}
%%%%%%%%%%%%%%%%%%%%%%%%%%%%%%%%%%%%%%%%%%%%%%%%%%%%%%%%%%%%%%%%%%%%%%%%%%%%%%%%
We summarize here the asymptotic formulas of the MFPT \ref{eq} when the domain $\Omega$ is in the plane and absorbing boundary is a small sub arc $\p\Omega_a$ (of length $a$) of the boundary $\p\Omega$. We have reviewed the mathematical method and the analysis in \cite{SIAM2014,RPP2013}. There is little intuition behind these formulas and it is not fruitful to guess what there are. It is indeed hard to tell in advance how the geometry enters into the formulas.  The recipe we adopted is to follow the analytical derivations that reveal how local and global structures, smoothness or not, local curvature controls the narrow escape time.

\begin{enumerate}%[(i)]
\item When $\p\Omega_a$ is a sub-arc  of a smooth boundary, the MFPT from any point $\x$ in $\Omega$ to $\p\Omega_a$ is denoted
$\bar\tau_{\small\x\to\p\Omega_a}$. For
{
 \beq
 \eps=\frac{\pi|\p\Omega_a|}{|\p\Omega|}=\frac{\pi a}{|\p\Omega|}\ll1\label{eps}
 \eeq
the MFPT is independent of $\x$ outside a small vicinity of $\p\Omega_a$ (called a boundary layer). Thus for
$\x\in\Omega$, outside a boundary layer near $\p\Omega_a$,
\beq
\bar\tau_{\small\x\to\p\Omega_a}=\frac{|\Omega|}{\pi D}\ln
 \frac{1}{\eps}+O(1),\label{taudisk}
\eeq
} where $O(1)$ depends on the initial distribution of $\x$ \cite{Ward1}--\cite{PNAS}. This result was derived independently using matched asymptotic technics and Green's function method.

If $\Omega$ is a disc of radius $R$, then for $\x$ at the center of the disk (Fig. \ref{f:drawing-classificationF}A),
 \beqq% \label{main:mfpt-center}
 \bar\tau_{\small\x\to\p\Omega_a}
=\frac{R^2}D\left[\log\frac{R}{a} + 2\log 2 + \frac{1}{4} +
O(\varepsilon)\right],
 \eeqq
and averaging with respect to a uniform distribution of $\x$ in the disk \cite{HolcmanSchuss2015}
 \beqq
%\label{main:averaged}
\bar\tau= \frac{R^2}D\left[\log\frac{R}{a} + 2\log 2 + \frac{1}{8} +
O(\varepsilon)\right].
 \eeqq
This result was obtained from generalizing Sneddon's method for mixed boundary value problem. The method is based on Abel's transformation \cite{NarrowEscape1}-\cite{NarrowEscape3}. The flux through a hole in a smooth wall on a flat membrane surface is regulated by the area $|\Omega|$ inside the wall, the diffusion coefficient $D$, and the aspect ratio $\eps$ (\ref{eps}). In the case of Brownian motion on a sphere of radius $R$ the MFPT to an absorbing circle centered on the north-south axis near the south pole with small radius
$a=R\sin\delta/2$  is given by
\begin{equation}
\label{eq:v-sphere} \bar\tau= \frac{2R^2}{D} \log\frac{\sin \frac \theta2}{\sin\frac\delta2},
\end{equation}
where $\theta$ is the angle between $\x$ and the south-north axis of the sphere (Fig. \ref{f:drawing-classificationF}B).
\item If the absorbing window is located at a corner of angle $\alpha$, then
\beq
\bar\tau = \ds{
\frac{|\Omega|_g}{D\alpha}}\left[\log\ds{\frac{1}{\varepsilon}}+O(1)\right],\label{log}
\eeq
where $|\Omega|_g$ is the {surface area of the domain on the curved surface, calculated according to the Riemannian metric on the surface} \cite{NarrowEscape1}. {Formula (\ref{log}) indicates that control of flux is regulated also by the access to the absorbing window afforded by the angle of the corner leading to the window (Fig. \ref{f:drawing-classificationF}C). This formula was obtained using a conformal map sending a corner to a flat line.}
\item If the absorbing window is located at a cusp, then $\bar\tau$ grows algebraically, rather than logarithmically. Thus, in the domain bounded between two tangent circles, the expected lifetime is
 \beq
\bar\tau= \ds{\frac{|\Omega|}{(d^{-1}-1)D}}\left(\frac{1}{\varepsilon} + O(1)\right),\label{taucusp}
 \eeq
where $d<1$ is the ratio of the radii \cite{NarrowEscape3} (Fig. \ref{f:drawing-classificationF}F). {Formula (\ref{taucusp}) indicates that a drastic reduction of flux can be achieved by putting an obstacle that limits the access to the absorbing window by forming a cusp-like passage. This formula was derived using the exponential conformal map.}
\item When $\p\Omega_a$  (of length $a$) is located at the end of a narrow neck with radius
of curvature $R_c$, the MFPT is given in \cite{holcmanMMS2012,HolcmanSchuss2015} as (Fig. \ref{f:drawing-classificationF}G and I)
 \beq
\bar\tau=\frac{|\Omega|}{4D\sqrt{2a/R_c}}\left(1+O(1)\right)\hspace{0.5em}\mbox{for}\
a\ll|\p\Omega|.\label{bartau}
 \eeq
This formula is derived by a new method that uses a Mobius transformation to resolve the cusp singularity \cite{holcmanMMS2012,HolcmanSchuss2015}. The boundary layer at the cusp is sent to a banana shaped domain. Asymptotic formula for a general cusp with an arbitrary power law are not known.

 For a surface of revolution generated by rotating the curve about its axis of symmetry \cite{holcmanMMS2012}, we use the representation of the generating curve
\beqq
y=r(x),\ \Lambda<x<0\hspace{0.5em}
\eeqq
where the $x$-axis is horizontal with $x=\Lambda$ at the absorbing end $\AB$. We assume that the parts of the curve that generate the funnel have the form
\beq
r(x)&=O(\sqrt{|x|})\hspace{0.5em}\mbox{near $x=0$}\nonumber\\
r(x)&=a+\frac{(x-\Lambda)^{1+\nu}}{\nu(1+\nu)\ell^{\nu}}(1+o(1))\hspace{0.5em}\mbox{for
$\nu>0$ near $x=\Lambda$},\label{rzL}
\eeq
where $a=\frac12\overline{\AB}=\eps/2$ is the radius of the gap, and the
constant $\ell$ has dimension of length. For $\nu=1$ the parameter $\ell$ is
the radius of curvature $R_c$ at $x=\Lambda$. The MFPT from the head to the
absorbing end $\AB$ is given by
\beq
\bar\tau\sim\frac{{\cal S}(\Lambda)}{2D}
\frac{\left(\ds\frac{\ell}{(1+\nu)a}\right)^{\nu/1+\nu}\nu^{1/1+\nu}}{\sin\ds\frac{\nu
\pi}{1+\nu}}, \label{u0SD0}
\eeq
where ${\cal S}$ is the entire unscaled area of the surface. In particular, for
$\nu=1$ the MFPT (\ref{u0SD0}) reduces to
\beq
\bar\tau\sim\frac{{\cal S}}{4D\sqrt{a/2\ell}}.\label{u0Rc}
\eeq
\item When a bulky head is connected to an essentially one-dimensional
strip (or cylinder) of small radius $a$ and length $L$, as is the case of a neuronal
spine membrane (Fig. \ref{f:drawing-classificationF}D). The connection of the head to the neck can be at an angle or by a smooth funnel. The boundary of the domain reflects Brownian trajectories and only the end of the
cylinder $\p\Omega_a$ absorbs them.  The domain $\Omega_1$ is connected to the cylinder at an
interface $\p\Omega_i$, which in this case is an interval $AB$. The MFPT from
$\x\in\Omega_1$ to $\p\Omega_a$ is given by
\beq
\ds{\bar\tau_{\small\x\to\p\Omega_a}=\bar\tau_{\small\x\to\p\Omega_i}+
\frac{L^2}{2D}+\frac{|\Omega_1|L}{|\p\Omega_a|D}}.\label{taubar}
\eeq
The flux dependence on the neck length is quite strong. This formula is derived using the additive property of the MFPT \cite{Dire-Part-I}.
\item A dumbbell-shaped domain (of type (VI)) consists of two compartments $\Omega_1$ and $\Omega_3$ and
a connecting neck $\Omega_2$ that is effectively one-dimensional (Fig. \ref{f:drawing-classificationF}J), or in a similar domain with a long neck. A Brownian
trajectory that hits the segment $\AB$ in the center of the neck $\Omega_2$ is equally likely to reach
either compartment before the other; thus $\AB$ is the stochastic separatrix (SS). Therefore the mean time
to traverse the neck from compartment $\Omega_1$ to compartment $\Omega_3$ is asymptotically twice
the MFPT $\bar\tau_{\small\Omega_1\to SS}$. Neglecting, as we may, the mean residence
time of a Brownian trajectory in $\Omega_2$ relative to that in $\Omega_1$ or in
$\Omega_3$ we can write the transition rates from $\Omega_1$ to the $\Omega_3$ and vv as
\beq
\lambda_{\Omega_1\to \Omega_3}=\frac{1}{2\bar\tau_{\small\Omega_1\to
SS}},\quad\lambda_{\Omega_3\to \Omega_1}=\frac{1}{2\bar\tau_{\small\Omega_3\to
SS}}.\label{rates}
\eeq
These rates can be found from explicit expressions for the flux into an absorbing window
\beq
\lambda_1\sim\frac{1}{\bar\tau},\label{lambda1}
\eeq
where $\bar\tau$ is given in (\ref{taubar}). Here $\bar\tau_{\small\x\to\p\Omega_i}$ is any one of the MFPTs given above,
depending on the geometry of $\Omega_1$ with $L$ half the length of the neck
and with $SS=\p\Omega_a$. The radii of curvature $R_{c,1}$ and $R_{c,3}$ at the
two funnels may be different in $\Omega_1$ and $\Omega_3$. The smallest
positive eigenvalue $\lambda$ of the Neumann problem for the Laplace equation
in the dumbbell is to leading order $\lambda=-(\lambda_{\Omega_1\to
\Omega_3}+\lambda_{\Omega_3\to\Omega_1})$. For example, if the solid dumbbell
consists of two general heads connected smoothly to the neck by funnels (see
(\ref{bartau1/2})), the two rates are given by
\beq
\frac{1}{\lambda_{\Omega_1\to
\Omega_3}}=&\sqrt{2}\left[\left(\frac{R_{c,1}}{a}\right)^{3/2}
\frac{|\Omega_1|}{R_{c,1}D}\right](1+o(1))+\frac{L^2}{4D}+\frac{|\Omega_1|L}{\pi a^2D}\nonumber\\
&\label{lambdas}\\
 \frac{1}{\lambda_{\Omega_3\to
\Omega_1}}=&\sqrt{2}\left[\left(\frac{R_{c,3}}{a}\right)^{3/2}\frac{|\Omega_3|}{R_{c,3}D}\right](1+o(1))+\frac{L^2}{4D}
+\frac{|\Omega_3|L}{\pi a^2 D} \nonumber
\eeq
(see \cite{Dire-Part-I}). Formulas (\ref{lambdas}) indicate that the unidirectional fluxes
between the two compartments of a dumbbell-shaped domain can be controlled by the area (or surface area) of
the two and by the type of obstacles to the access to the connecting neck. The equilibration rate in the
dumbbell, $\lambda$, is thus controlled by the geometry.
\item The mean time to escape through $N$ well-separated absorbing windows of lengths
$a_j$ at the ends of funnels with radii of curvature $\ell_j$, respectively, in
the boundary $\p\Omega$ of a planar domain $\Omega$ is given by
\beq\label{bartaui}
\bar\tau=\frac{\pi|\Omega|}{{2}D\sum_{j=1}^N\sqrt{a_j/\ell_j}}\left(1+o(1)\right)\hspace{0.5em}\mbox{for}\
a_j/\ell_j\ll|\p\Omega|.
\eeq
The probability to escape through window $i$ is given by
\beq
p_i=\frac{\sqrt{a_i/\ell_i}}{\sum_{j=1}^N\sqrt{a_j/\ell_j}}.\label{pi}
\eeq
{Formulas (\ref{bartaui}) and (\ref{pi}) are significant for diffusion in a network of
compartments connected by narrow passages (e.g., on a membrane strewn with obstacles). The dependence of the
MFPT $\bar\tau$ and of the transition probabilities $p_i$ on the local geometrical properties of the
compartments renders the effective diffusion tensor in the network position-dependent and can give rise to
anisotropic diffusion.}
\end{enumerate}

%%%%%%%%%%%%%%%%%%%%%%%%%%%%%%%%%%%%%%%%%%%%%%%%%%%%%%%%%%%%%%%%%%%%%%%%%%%%%%%%%%%%%%%
%\begin{figure}[http!]
%    \centering
%        \includegraphics[width=0.35\textwidth]{Obstacles.eps}
%    \caption{{\bf Mean first encounter time (MFET) }}
%    \label{fig:NET1}
%\end{figure}
%%%%%%%%%%%%%%%%%%%%%%%%%%%%%%%%%%%%%%%%%%%%%%%%%%%
%%%%%%%%%%%%%%%%%%%%%%%%%%%%%%%%%%%%
\subsection{Narrow escape formula in three-dimension}
%%%%%%%%%%%%%%%%%%%%%%%%%%%%%%%%%%%%
{We now summarize the Narrow Escape Time formula in three-dimensions. The methods are the same as in two dimensions: Matched asymptotic or Greens function and conformal mapping to resolve cusp singularities. Indeed, the axial symmetry allows reducing the three- to two- dimensions and thus to use of conformal transformations \cite{HolcmanSchuss2015}.}
\begin{enumerate}
\item The MFPT to a circular absorbing window
$\p\Omega_a$ of small radius $a$ centered at $\mb{0}$ on the boundary
$\p\Omega$ is given by \cite{leakage}
 \beq
\bar\tau_{\small\x\to\p\Omega_a}=\frac{|\Omega|}{4aD\left[1+\ds\frac{L(\mb{0})+
N(\mb{0})}{2\pi}\,a\log a+o(a\log a)\right]}\label{tau3D},
 \eeq
where $L(\mb{0})$ and $N(\mb{0})$ are the principal curvatures of the boundary
at the center of $\p\Omega_a$. This formula was derived using the second order expansion of the Neuman-Green's function at the pole \cite{leakage}.
\item The MFPT from the head of the solid of revolution, obtained by rotating the
symmetric domain about its axis of symmetry, to a small absorbing window $\p\Omega_a$ at the end of a
funnel (Fig. \ref{f:drawing-classificationF}H) is given by
\beq
\bar\tau=\frac{1}{\sqrt{2}}\left(\frac{R_c}{a}\right)^{3/2} \frac{|\Omega|}{R_c
D}(1+o(1))\hspace{0.5em}\mbox{for}\ a\ll R_c,\label{bartau1/2}
\eeq
where the $R_c$ is the radius of curvature of the rotated curve at the end of
the funnel \cite{Dire-Part-I}.
\item The MFPT from a point $\x$ in a bulky head $\Omega$ to an absorbing disk $\p\Omega_a$ of a small radius
$a$ at the end of a narrow neck of length $L$, connected to the head at an
interface $\p\Omega_i$ is given by the connection formula (\ref{taubar}). When
the cylindrical neck is attached to the head at a right angle the interface
$\p\Omega_i$ is a circular disk and $\bar\tau_{\small\x\to\p\Omega_i}$ is given
by (\ref{tau3D}). When the neck is attached smoothly through a funnel,
$\bar\tau_{\small\x\to\p\Omega_i}$ is given by (\ref{bartau1/2}).
\item The mean time to escape through $N$ well-separated absorbing circular windows
or radii $a_j$ at the ends of funnels with curvatures $\ell_j$, respectively,
is given by
\beq
\bar\tau=\frac{1}{\sqrt{2}}\,\frac{|\Omega|}{D\sum_{j=1}^N\ell_j\ds{\left(\frac{a_j}{\ell_j}\right)^{3/2}}}
\label{bartauN}.
\eeq
The exit probability through window $i$ is given by
\beq
p_i=\frac{{a_i}^{3/2}\ell_i^{-1/2}} {\sum_{j=1}^N{a_j}^{3/2}\ell_j^{-1/2}}.
\eeq
\item The principal eigenvalue of the Laplace equation in a dumbbell-shaped
structure is given in item (vi), equations (\ref{rates})--(\ref{lambdas}) above
\cite{Dire-Part-I}.
\item The leakage flux through a circular hole of small radius $a$ centered at $\mb{0}$ in the
reflecting boundary is given by \cite{leakage}
 \beq
J_{a}=4aDu_0(\mb{0})+O\left(\frac{a^2}{|\Omega|^{2/3}}\log\frac{a}{|\Omega|^{1/3}}\right),\label{Intro-Jeps4}
 \eeq
where $u_0(\mb{0})$ is the concentration of diffusers at the window in the same model
without the absorbing window.
\end{enumerate}
We refer the reader to the classical literature about the asymptotic formula for Narrow Escape Time \cite{PhysA2014}, the dire Strait time (when escape occurs at a cusp boundary)\cite{holcmanMMS2012} and the recent monograph \cite{HolcmanSchuss2015} for applications in cellular biology.

%%%%%%%%%%%%%%%%%%%%%%%%%%%%%%%%%%%%%%%%%%
\section{Stochastic Smoluchowski equation for modeling polymer dynamics}
%%%%%%%%%%%%%%%%%%%%%%%%%%%%%%%%%%%%%%%%%%
A significant application of the Smoluchowski's limit equation is to polymer models. The Rouse model is defined as a collection of beads connected by springs \cite{Doi}. Monomers are positioned at $\vect R_n$ ($n=1,2,...N$), subject to Brownian motions and the spring forces are due to the coupling between the nearest neighboring beads. The potential energy is defined by
\beq\label{RousePot}
\phi(\vect R) = \frac{\kappa}{2}\sum_{n=1}^N \left(\vect R_n - \vect R_{n-1} \right)^2,
\eeq
In the Rouse model, only neighboring monomers interact \cite{Doi}. In the Smoluchowski's limit of the Langevin equation, the dynamics of monomer $\vect R_{n}$ is driven by the potential $\phi(\vect R_1,..,\vect R_N)$, which generates the force $-\nabla_{\vect R_n}\phi(\vect R_1,..,\vect R_N)$. The ensemble of stochastic equations is
\beq\label{RouseEq}
\frac{d \vect{ R_n}}{d t} = -D\kappa (2\vect R_n - \vect R_{n-1} - \vect R_{n+1})+ \sqrt{2D}\frac{d\vect{w_n}}{dt}
\eeq
for $n=1,..N$. In this model, at equilibrium all beads are centered at zero, but the variance of the distances in a polymer realization is given by
\beq
<|R_{n+1}-R_{n}|^2>=b^2,
\eeq
where $b$ is the standard deviation of the bond length, $\kappa=d k_BT/b^2$ is the spring constant with $d$ the spatial dimension, $k_B$ is the Boltzmann coefficient and $T$ the temperature. For a freely-joint-chain polymer, the energy between monomer is changed to
\beq\label{RousePot2}
\phi(\vect R_1,..,\vect R_N) = \frac{\kappa}{2}\sum_{n=1}^N \left(|\vect R_n - \vect R_{n-1}|-l_0 \right)^2,
\eeq
leading to a steady state configuration, where the mean distance between neighboring beads is $<|R_{n+1}-R_{n}|>=l_0$, where by taking the limit $l_0=0$, we recover the classical Rouse model. Starting with a given configuration, the relaxation of a Rouse polymer to steady state in a free space can be analyzed using the Fourier space
\beq\label{DCT}
\vect{u_p} = \sum^{N}_{n=1} \vect R_n \alpha^n_p,
\eeq
where the change of coordinates is encoded in the matrix
\begin{equation}\label{Rousecoef}
\alpha^n_{p}=\left\{
\begin{array}{cc}
\sqrt{\frac{1}{N}},\ &\textrm{} p=0 \\
\sqrt{\frac{2}{N}} \cos \left((n-1/2) \frac{p\pi}{N}\right), &\textrm{ otherwise}.
\end{array}%
\right.
\end{equation}
$\vect u_0$ represents the motion of the center of mass and the potential $\phi$ defined in equation \ref{RousePot} is now
\beq\label{Rouse_pot3}
\phi (\vect{u_1},..,\vect{u_{N-1}})= \frac{1}{2} \sum^{N-1}_{p=1} \kappa_p \vect u^2_p,
\eeq
where
\beq\label{Rouse_eigen}
\kappa_p = 4\kappa \sin( \frac{p \pi}{2N})^2.
\eeq
Equations \ref{RouseEq} are now decoupled in a $(N-1)d-$independent Ornstein-Uhlenbeck (OU) processes
\beq\label{RouseModeEq}
\frac{d \vect u_p}{d t} = -D_p \kappa_p \vect u_p  + \sqrt{2D_p}\frac{ d\vect{\widetilde{w_p}}}{dt},
\eeq
where $\vect{\widetilde{w_p}}$ are independent $d$-dimensional Brownian motions with mean zero and variance $1$ and $D_p=D$ for $p=1..N-1$, while $D_0=D/N$ and the relaxation times are defined by $\tau_p=1/D\kappa_p$. The center of mass behaves as a freely diffusing particle. Starting from a straight line, the time to relax for a Rouse polymer is dominated by the slowest time constant
\beq\label{RouseRM}
\tau_N=\frac1{D\kappa_1}=\frac1{4D\kappa \sin (\frac{1\pi}{2N})^2}\approx \frac{N^2}{D\kappa \pi^2}.
\eeq

\subsection{Anomalous motion of a Rouse polymer}\label{s:anomalous}
%%%%%%%%%%%%%%%%%%%%%%%%%%%%%%%%%%%%%%%%%%%%%%%%%%%%
The motion of monomer $\vect R_c$ of a Rouse polymer is related to the Fourier coefficients by
\beq\label{eq:def_inv}
\vect R_c = \sum_{p=0}^{N-1} \alpha^c_p \vect u_p,
\eeq
where $\alpha^c_p$ are described by relation \ref{Rousecoef} and $\vect u_p$ satisfy eqs.\ref{RouseModeEq}, which form an ensemble of Ornstein-Uhlenbeck processes, for which the variance is simply
\beq\label{eq:sigma_p}
\begin{aligned}\label{varianceRP}
\sigma_p^2(t) &=\langle|\vect u_p(t) -\vect u_p(0)|^2\rangle =\frac{1}{\kappa_p}(1-{\rm e}^{-2t/\tau_p}),\ {\rm for\ }p\ge 1\,, \\
\sigma_0^2(t) &= 2D_{\textrm{cm}}t\,.
\end{aligned}
\eeq
The relaxation times are defined by
\beq\label{relaxation_times}
\tau_p=\frac{1}{D\kappa_p},
\eeq
while the diffusion constant is $D_{\textrm{cm}}=D/N$. The shortest timescale is $\tau_{N-1}\approx 1/(4D\kappa)$ which is half of the time $\tau_{\rm s}=1/(2D\kappa)$ for a free monomer to diffuse a mean squared distance between adjacent monomers ($b^2=1/\kappa$). The center of mass is characterized by the time scale $\tau_0\equiv b^2N/D_{\rm{cm}}=N^2/(D\kappa)$ which is the time for a particle to diffusion across the polymer size. For long polymers $\tau_0/\tau_1\approx\pi^2$. Using relation \ref{varianceRP}, the MSD of  monomer $\vect R_c$ is a sum of independent OU-variables,
\beq\label{correlationRP}
{\rm var}(\vect{R_c}(t))=\langle ( \vect R_c(t) - \vect R_c(0))^2 \rangle = \frac{d}{\kappa N} \sum^{N-1}_{p=1} \frac{ \cos^2 \left( \frac{(2c-1)p\pi}{2N}\right)}{\sin^{2}(\frac{p\pi}{2N})}(1 - e^{-2t/\tau_p}) + 2 d D_{\rm cm}t,
\eeq
where $d$ is the spatial dimension. Formula \ref{correlationRP} shows the deviation with MSD of a Brownian motion, for which the correlation function increases linearly with time. There are three distinguish regimes:
%%%%%%%%%%%%%%%%%%%%%%%%%%%%%%%%%%%%%%%%%%%%%%%%%%%
\begin{enumerate}
\item For short time $t\ll \tau_{N-1}$, $\sigma_p^2(t)\approx Dt$, independent of $p$, the sum in eq.~\eqref{correlationRP} leads to
\beq
{\rm var}(\vect{R_c})\approx 2dDt,
\eeq
which is the diffusion regime.
\item For large time, $t\gg \tau_1$, the exponential terms in relation \ref{correlationRP} becomes independent of $t$. Only the first term in Eq.~\eqref{varianceRP} corresponding to the diffusion of the center of mass gives the time-dependent behavior. This regime is dominated by normal diffusion, with a diffusion coefficient $D/N$.
\item For intermediate times $\tau_{N-1}\ll t\ll \tau_1$, such that $2t/\tau_{p}>1$, the sum of exponentials contributes to eq.\eqref{correlationRP}. The variance \ref{correlationRP} is
\beq\label{eq:approxVar_R_c}
{\rm var}(R_c)\approx \ds{2\int_{p_{\rm min}}^{N-1}\frac{ \cos^2 \left( \frac{(2c-1)p\pi}{2N}\right)}{\sin^{2}(\frac{p\pi}{2N})}}dp,
\eeq
where $p_{\rm min}$ is such that $\tau_{p_{min}}=2t$. We have ${\rm var}(\vect R_c)\sim t^{1/2}$. A Rouse monomer exhibits anomalous diffusion.  The time interval can be arbitrarily long with the size $N$ of the polymer.
\end{enumerate}
%%%%%%%%%%%%%%%%%%%%%%%%%%%%%%%%%%%%%%%%%%%%%%%%%%%%%%%%%%%5

%%%%%%%%%%%%%%%%%%%%%%%%%%%%%%%%%%%%%
\subsection{Looping time: a brief summary of an analytical approach}
%%%%%%%%%%%%%%%%%%%%%%%%%%%%%%%%%%%%%
The first looping time between two monomers is the First Encounter Time $ \tau_e$ for two monomers $n_a,n_b$ to come into a distance $\varepsilon<b$, defined by
\beq\label{contact_mon}
\tau_e =\inf\{ t>0 \hbox{ such that }|\vect R_{n_a}(t) - \vect R_{n_b}(t)| \leq \varepsilon\},
\eeq
where $\vect R_{n_a}$ and $\vect R_{n_b}$ follows for example the Rouse equation \ref{RouseEq}. We present the asymptotic computation of the Mean First Encounter Time (MFET) $\langle \tau_e \rangle$ for the two ends $\vect R_N,\vect R_1$ meets. The two monomers meet when distance is less than $\varepsilon<b$, that is
\beq\label{eq:loop_real}
\left| \vect R_N - \vect R_1 \right| \leq \varepsilon
\eeq
In Rouse coordinates, $ \vect{u_p} = \sum^{N}_{n=1} \alpha^n_p \vect R_n$
where $\alpha^n_{p}$ are defined in \ref{Rousecoef}, condition \eqref{eq:loop_real} is
\beq\label{S_boundary_condition}
\left| 2\sqrt{\frac{2}{N}} \ds \sum_{p \; \rm{odd}} \vect u_p \cos(p\pi/2N)
\right|\leq \varepsilon.
\eeq
The end-to-end encounter is independent of the center of mass, with coordinate $\vect u_0$. Thus, the MFET is the MFPT for the $(N-1)d$-dimensional stochastic process
\beq
\vect u(t) =(\vect u_1(t),..\vect u_{N-1}(t))
\in \Omega \times \Omega...\times \Omega=\tilde \Omega,
\eeq
where $\Omega=\rR^2$ or $\rR^3$ and $\vect u_p$ satisfies the OU-equations \ref{RouseModeEq}
%\beq\label{eq:eta_dynamics}
%\frac{d \vect u_p}{d t} = -D_p \kappa_p \vect u_p  + \sqrt{2D_p}\frac{ d\vect{  \widetilde{w_p}}}{dt},
%\eeq
%($D_p=D,\kappa_p=4\kappa \sin\(p\pi/2N\)^2$ and $p=1,..,N-1$)
to the boundary of the domain
\beq \label{seps}
 S_{\epsilon}=\{P \in \tilde \Omega \hbox{ such that } dist(P,{\cal S})\leq
\frac{\varepsilon}{\sqrt{2}} \},
\eeq
%Each $\vect{\widetilde{w_p}}$ is independent $d$-dimensional Brownian motions with mean zero and variance $1$,
where $dist$ is the Euclidean distance and
\beq\label{defS}
{\cal S}=\{{(\vect u_1,..\vect u_{N-1})  \in \tilde \Omega } \big |\sum_{p \; \rm{odd}} \vect u_p \cos(p\pi/2N) =0\}
\eeq
is a submanifold of codimension $d$ in $\tilde \Omega $. The probability density function (pdf) $p(\vect u(t)=\vect x,t)$ satisfies the forward Fokker-Planck equation (FPE) \cite{Schuss2}
\beq\label{FFP2}
\frac{1}{D}\frac{\partial p(\vect x,t)}{\partial t}&=& \Delta p(\vect x,t) + \nabla\cdot\(  \nabla\phi \:p(\vect x,t)\) = \mathcal{L} p,\nonumber\\
p(\vect x,0)&=&p_0(\vect x), \label{abs_cond}
\eeq
with boundary condition $ p(\vect x,t)=0 $ for $x\in \partial S_{\epsilon}$, $p_0(\vect x)$ is the initial distribution and the potential $\phi(u_1,..,u_N) = \frac{1}{2}\sum_{p} \kappa_p \vect u^2_p$ was introduced in \ref{Rouse_pot3}.

The solution of equation \eqref{FFP2} is expanded in eigenfunctions
\beq\label{eq:eigen_F_FP}
p(\vect x,t) = \ds \sum^{\infty}_{i=0} a_i
w_{\lambda^{\epsilon}_i}( \vect x) e^{-\lambda^{\epsilon}_i tD}
e^{-\phi\left(\vect x
\right)},
\eeq
where $a_i$ are coefficients, $w_{\lambda^{\epsilon}_i}( \vect x)$ and $\lambda^{\epsilon}_i $ are the eigenfunctions and eigenvalues respectively of the operator $\mathcal L$ in the domain $\Omega_{\epsilon} = \tilde \Omega-S_{\epsilon}$. The probability distribution that the two ends have not met before time $t$ is the survival probability
\beq
p(t) = Pr\{ \tau_{\epsilon}>t\}=\int_{\Omega_{\epsilon} }p(\vect x,t)dx,
\eeq
and the first looping time is
\beq
\tau_\epsilon =\inf\{t>0, \vect u(t) \in \p S_{\epsilon}\}.
 \eeq
Using expansion \ref{eq:eigen_F_FP},
$ p(t)=4 \sum^{\infty}_{i=0} C_i e^{-\lambda^{\epsilon}_i D t}$ where $C_i= \int_{\Omega_{\epsilon}} p_0(\vect x) w_{\lambda^{\epsilon}_i} (\vect x)d \vect x \int_{\Omega_{\epsilon}} w_{\lambda^{\epsilon}_i} (\vect x) e^{-\phi\left(\vect x \right)}d \vect x$. Starting with an equilibrium distribution $p_0(\vect x) = |\tilde \Omega|^{-1} e^{-\phi\left(\vect x \right)}$, we have
\beqq
C_i =|\tilde \Omega|^{-1} (\int_{\Omega_{\epsilon}} w_{\lambda^{\epsilon}_i} (\vect x) e^{-\phi\left(\vect x \right)}d\vect x)^2
\eeqq
and finally the MFET is given by
\beq\label{MFLT_eigen}
\langle \tau_\epsilon \rangle =\sum^{\infty}_{i=0} \frac{C_i}{D\lambda^{\epsilon}_i}.
\eeq
Starting from the equilibrium distribution $C_0\approx 1$, while the other terms are $C_i=o(1)$, as we shall see, the first term is the main contributor of the series.

%%%%%%%%%%%%%%%%%%%%%%%%%%%%%%%%%%%%%%%%%%%%%
\subsection{Computing the eigenvalues of the Fokker-Planck equation and the MFET} \label{s:computingMEST}
%%%%%%%%%%%%%%%%%%%%%%%%%%%%%%%%%%%%%%%%%%%%%
The eigenvalues $\lambda^{\epsilon}_i$ of the operator $\mathcal{L}$ (eq.\ref{FFP2}) are obtained by solving the forward FPE in $\rR^{d(N-1)}$, with the zero absorbing boundary condition on the entire boundary of the domain $S_{\epsilon}$ (see eq. \ref{seps}), which is the tubular neighborhood of the $(N-1)d$-dimensional sub-manifold ${\cal S}$. For small $\varepsilon$, the eigenvalues can be computed from the following regular expansion near the solution of the domain with no domain $ S_{\epsilon}$:
\beq\label{chavel}
\lambda^{\epsilon}_i&=&\lambda^{0}_{i} +c_2 \epsilon \int_{S} w_{\lambda^{0}_{i}}^2 dV_x+\mathcal O(\epsilon^2), \hbox{ for } d=3 \label{chavel3d}\\
\lambda^{\epsilon}_i&=&\lambda^{0}_{i} + \frac{2\pi}{ \log \epsilon} \int_{S} w_{\lambda^{0}_i}^2 dV_x+\mathcal O\(\(\frac{1}{\log \epsilon}\)^2\) \hbox{ for } d=2, \label{chavel2d}
\eeq
where the eigenfunction $w_{\lambda^{0}_i}$ and eigenvalues $\lambda^0_{i}$ are associated to the non perturbed operator (no boundary) \cite{Chavel1988},  $d=3,2$.

In the context of the Rouse polymer, the volume element is $dV_x = e^{-\phi(\vect x)} d\vect x_g$, $d\vect x_g$, a measure over the sub-manifold ${\cal S}$ and $c_2=\frac{2\pi^{3/2}}{\Gamma(3/2)}$ \cite{Chavel1988}.  The unperturbed eigenfunctions $w_{\lambda^{0}_i}$ are products of Hermite polynomials \cite{Abramowitz:Book}, that depend on the spatial coordinates and the eigenvalues $\lambda^{0}_i$ are the sum of one dimensional eigenvalues \cite{Amitai2012_1}. The first eigenfunction associated to the zero eigenvalue is $w_{\lambda^{0}_{0}} = |\tilde \Omega|^{-1/2}$.  The first eigenvalue for $\varepsilon$ small is obtained from relation \ref{chavel3d} in dimension 3 with $\lambda^{0}_{i} =0$,
\beq\label{firsteigen3d}
\lambda^{\epsilon}_0 = \frac{c_2 \epsilon \ds \int_{S} e^{-\phi\left(\vect x \right) }
 d\vect x_g }{|\tilde \Omega|}+ \mathcal{O(}\epsilon^2 \mathcal{)},
\eeq
which is the ratio of the closed to all polymer configurations. Using the potential $\phi$ (defined in \ref{Rouse_pot3}), the volume is computed explicitly  from Gaussian integrals
\beq
|\tilde \Omega| = \int_{\Omega} e^{-\phi\left(\vect x \right) }d\vect x_g = \left[\frac{(2\pi)^{(N-1)}}{\prod_1^{N-1} \kappa_p} \right]^{d/2},
\eeq
while the parametrization of the constraint \ref{defS} leads to
\beq
\int_{S} e^{-\phi \left(\vect x \right) }d\vect x_g  = \left[\frac{(2\pi)^{N-2} \prod_{p \;
\rm{odd}}\omega^2_p}{\prod_{p}\kappa_p\( \sum_{p \; \rm{odd}} \frac{\omega^2_p}{\kappa_p} \)} \right]^{d/2},
 \eeq
where $\omega_p = \cos(p\pi/2N)$.  In summary, for fix $N$ and small $\varepsilon$,
\beq\label{first_eigen_eps}
  \lambda^{\epsilon}_{0} = \left\{
  \begin{array}{l l}
     \ds \left( \frac{\kappa}{N\pi}\right)^{3/2}4\pi \epsilon + \mathcal{O}(\epsilon^2) & \quad \text{for } d=3, \\\\
    \ds \frac{2\kappa}{N\log\left(\frac{\sqrt{2}b}{\varepsilon}\right)} + \mathcal{O}\(\(\frac1{\log\epsilon}\)^2\) & \quad \text{for } d=2.\\
  \end{array} \right.
\eeq
The zero eigenvalue is sufficient to characterize the MFET, confirming that the FET is almost Poissonian, except for very short time. Moreover, the second term in the expansion of $\lambda^{\epsilon}_{0}$ is proportional to $1/N$. Using the approximation $C_0 \sim1$ and relation \ref{MFLT_eigen}, for $d=3$, the MFET is approximated by
\beq
\langle \tau_\varepsilon \rangle_{3d}&\approx&\frac{1}{ D\lambda^{\epsilon}_{0} } =
\frac{1}{D\( \left( \frac{\kappa}{N\pi}\right)^{3/2}4\pi \epsilon -A\epsilon^2/N\)},
\eeq
where $A$ is a constant that has been estimated numerically. Indeed, with $\epsilon=\frac{\varepsilon}{\sqrt{2}}$, the MFET is for d=3
\beq\label{MFLT3d}
\langle \tau_\varepsilon \rangle_{3d} &=& \left( \frac{N\pi}{\kappa}\right)^{3/2}\frac{\sqrt{2}}{D 4\pi \varepsilon} + A_3 \frac{b^2}{D}  N^2+\mathcal O(1),
\eeq
which hold for a large range of $N$, as evaluated with Brownian simulations (Fig. \ref{fig:fitting_MFLT}). The value of the coefficient is $A_3=0.053$ \cite{Amitai2012_1} ($A_3=0.053b^2/D$). These estimates are obtained for fixed $N$ and small $\eps$.

Similar for $d=2$, a a two dimensional space, the asymptotic formula for MFET \cite{Amitai2012_1} is
\beq \label{MFLT2d}
\langle \tau_\varepsilon \rangle_{2d} &=& \frac{N}{2D\kappa}\log\left(\frac{\sqrt{2}b}{\varepsilon}\right) + A_2 \frac{b^2}{D} N^2+\mathcal O(1),
\eeq
All these asymptotic expansions are derived for fix $N$ and small $\eps$. However, there should not be valid in the limit $N$ large, although stochastic simulations (Fig. \ref{fig:fitting_MFLT}a-b) shows that the range validity is broader than expected. The exact asymptotic formula for any two monomers inside a polymer chain should be derived. Other scaling laws have been derived in \cite{Weiss}.

%%%%%%%%%%%%%%%%%%%%%%%%%%%%%%%%%%%%%%%%%%%%%%%%%%%%%%%%%%%%%%%%%%%%%%%%%%%%%%
\section{Diffusion approximation by jump processes and model of a membrane crowded with obstacles}
%%%%%%%%%%%%%%%%%%%%%%%%%%%%%%%%%%%%%%%%%%%%%%%%%%%%%%%%%%%%%%%%%%%%%%%%%%%%%%
 Smoluchowski equation has also been used to analyse diffusion in crystal or crowded medium such as cellular membrane. The approach consists of approximating diffusion jump process as continuous diffusion, valid at a much coarser-time scale than the continuous process itself. This approximation allows deriving asymptotic formula and interpreting data \cite{PNAS}. The transition between diffusion at a molecular level and sub-cellular level correspond to changing scale and is obtained by coarse-graining a model of disk obstacles using the narrow escape theory into a Markov process, which is a continuum approximation of the diffusion equation. The transition between the molecular and cellular regime occurs at a time scale characterized by the NET and is often interpreted as anomalous diffusion.

%%%%%%%%%%%%%%%%%%%%%%%%%%%%%%%%%%%%%%%%%%%%%%%%%%%%%%%%%%%%%%%%%%%%%%%%%%%%%%
The organization of a cellular membrane is to a large extent the determinant of
the efficiency of molecular trafficking of receptors to their destination.
The arrival rates of these molecules at their specific destinations control
their role and performance, and thus steer the cell toward its function.
After two decades of intense research on membrane organization, it is still unclear how the heterogeneity of the membrane controls diffusion (see Figure {\ref{f:Membrane-filaments}}).
Recently, using single molecule tracking, the diffusion coefficient of a
molecule freely diffusing on intact and treated neuronal membranes, cleared of
almost all obstacles was found. In this case the diffusion of a protein
on the membrane is described by the \index{Saffman-Delbr\"uck theory|textit} Saffman-Delbr\"uck theory. If,
however, the membrane is crowded with obstacles, such as fixed proteins, fences and pickets, and so
on, the effective diffusion coefficient differs significantly from that predicted in and
depends strongly on the degree of crowding. The latter can be estimated from diffusion data and from an appropriate model and its analysis, as explained below. The key
to assessing the crowding is to estimate the local diffusion coefficient from
the measured molecular trajectories and the analytic formula for the MFPT
through a narrow passage between obstacles.

%%%%%%%%%%%%%%%%%%%%%%%%%%%%%%%%%%%%%%%%%%%%%%%%%%%%%%%5
\subsection[Coarse-grained model of membrane crowding]{A coarse-grained model of membrane crowding organization}\label{ss:Coarse-grained}
%%%%%%%%%%%%%%%%%%%%%%%%%%%%%%%%%%%%%%%%%%%%%%%%%%%%%%%5
A simplified model of a crowded membrane can be a square lattice
of circular obstacles of radius $a$ centered at the corners of lattice squares
of side $L$ (Figure {\ref{f:figureobstacles_new}). The mean exit time from a
lattice box, formula (\ref{bartau}), is to leading order independent of the
starting position $(x,y)$  and can be approximated  as
\begin{align}\label{bartaun}
\bar\tau_4=\frac{\bar\tau}{4},
\end{align}where $\bar\tau$ is the MFPT to a single absorbing window in a narrow strait
with the other windows closed (reflecting instead of absorbing). It follows
that the waiting time in the cell enclosed by the obstacles is exponentially
distributed with rate
\begin{align}\lambda=\frac{1}{\bar 2\tau_4},\label{lambda}
\end{align}
where $\bar\tau$ is given by (\ref{log}) and (\ref{bartau}) as
\begin{align} \label{asympto2}
 \bar \tau \approx \left\{\begin{array}{ll}
&c_1 \mbox{ for } 0.8<\eps<1,  \\
&\\
&c_2|\Omega|\log{\ds\frac{1}{\eps}}+d_1 \mbox{ for } 0.55<\eps<0.8,  \\
&\\
&c_3\ds\frac{|\Omega|}{\sqrt{\eps}}+d_2 \mbox{ for } \eps<0.55,
\end{array}     \right.
\end{align}
with $\eps=(L-2a)/a$ and $d_1,d_2=O(1)$ for $\eps\ll1$  (see Figure \ref{f:figureobstacles_new}).
The MFPT $c_1$ from the center to the boundary of an unrestricted square is computed from
%\begin{widetext}
 \begin{align}
u(x,y) = \frac{4L^2}{\pi^3D} \sum_{0}^{\infty}
\ds{\frac{\left[\cosh(k+\frac12)\pi-\cosh(k+\frac12)\pi(2y/L-1)\right]\sin(2k+1)\pi
x/L} {(2k+1)^3\cosh(2k+1)\pi}},\label{separation}
 \end{align}
% \end{widetext}
so $c_1=u(L/2,L/2)\approx [4L^2/\pi^3D][\cosh(\pi/2-1)/\cosh\pi].$ For
$L=1,D=1$, we find $c_1\approx 0.076\,$, in agreement with Brownian dynamics
simulations (Fig. \ref{f:figureobstacles_new}B). The coefficient $c_2$ is obtained from \eqref{taudisk} as $c_2=1/2\pi D\approx 0.16.$ Similarly, the coefficient $c_3$
is obtained from \eqref{bartau} as $ c_3\approx \pi/4\sqrt{2}\,D\approx 0.56.$ The coefficients $d_i$ are chosen by patching $\bar\tau$ continuously between the different regimes:
\begin{align}
d_1 =c_1+ c_2|\Omega(r_1)|\log(1-2r_1),
\end{align}
and
 \beqq
 d_2= c_1-c_2 \left[|\Omega(r_1)|\log(1-2r_1)+|\Omega(r_2)|\log(1-2r_2)\right]\\
-c_3|\Omega(r_2)|(1-2r_2)^{-1/2},
\eeqq
where $|\Omega(r)|=L^2-\pi r^2$.

Simulations with $D=1$ in a square of radius $L=1$ with four reflecting circles of radius $r$, centered at the corners,
show that the uniform approximation by the patched formula (\ref{asympto2}) is
in good agreement with Brownian results (Fig. \ref{f:figureobstacles_new}b), where the
statistics were collected from 1,000 escape times of Brownian trajectories per
graph point. The trajectories start at the square center. Equation (\ref{asympto2}) holds in the full range of values of $a \in [0,L/2]$ and all $L$.\\

The Brownian motion around the obstacles (Figure {\ref{f:figureobstacles_new}(a)}) can be coarse-grained into a
\index{Markov jump process|textit} Markovian jump process whose state are the connected domains enclosed by the obstacles and the jump rates
are determined from the reciprocals of  the mean first passage times and exit probabilities. This \index{random walk|textit}  random walk can in turn be approximated by an effective coarse-grained anisotropic diffusion. The diffusion approximation to the transition
probability density function of an isotropic random walk that jumps at
exponentially distributed waiting times with rate $\lambda$ on a square lattice
with step size $L$ is given by \cite{Schuss2}
\begin{align}\frac{\p p}{\p t}=\bar D\left(\frac{\p^2p}{\p x^2}+\frac{\p^2p}{\p
y^2}\right),\quad \bar D=\frac{\lambda L^2}{4}.\label{barD}
\end{align}

%%%%%%%%%%%%%%%%%%%%%%%%%%%%%%%%%%%%%%%%%%%%%%%%%%%%%%%%%%%%
\subsection{ Diffusion of receptors on the \index{neuronal membrane|textit} neuronal membrane}\label{s:diffusionreceptor}
%%%%%%%%%%%%%%%%%%%%%%%%%%%%%%%%%%%%%%%%%%%%%%%%%%%%%%%%%%%%
The results of the previous section can be used to estimate the density of obstacles
on the membrane of cell such as a neuronal dendrite. The effective diffusion coefficient of a
receptor on the neuronal membrane can be estimated from the experimentally measured
single receptor trajectory by a single particle tracking method. The receptor effective diffusion
coefficient of a receptor varies between 0.01 and 0.2 $\mu$m$^2$/sec.

In the simplified model of crowding, the circular obstacles are as in (Fig.
\ref{f:figureobstacles_new}). Simulated Brownian trajectories give the MFPT from one square to the
 next one as shown in Fig. \ref{f:Membrane-filaments}, where $L$ is fixed and $a$ is variable.
According to \eqref{asympto2}, \eqref{lambda}, and \eqref{barD}, as $a$ increases the effective
diffusion coefficient $\bar D$ decreases. It is computed as the as the mean square displacement
(MSD) $\langle \frac{MSD(t)}{4t} \rangle$. Brownian simulations show that $\bar D$ is linear, thus
confirming that in the given geometry crowding does not affect the nature of the Brownian motion
for sufficiently long times. Specifically, for Brownian diffusion coefficient $D=0.2\,\mu$m$^2/s$
the time considered is longer than 10 $s$. In addition, Fig. \ref{f:fig5}c shows the
diffusion coefficient ratio $D_a/D_0$, where $D_a$ is the effective diffusion coefficient of
Brownian motion on the square lattice described above with obstacles of radius $a$. For $a=0.3$ the
value $D_a/D_0\approx 0.7$ is found whereas a direct computation from the mean exit time formula
\eqref{asympto2} gives
\begin{align}
\frac{\tau_0}{\tau_a} =\frac{c_1}{c_2|\Omega|\log \ds\frac{1}{\eps}+d_1}\approx 0.69,
\end{align}
where $\eps= (L-2a)/L=0.4$.

It can be concluded from the Brownian simulations that the coarse-grained motion is plain diffusion
with effective diffusion coefficient $D_a/D_0=\tau_0/\tau_a$, which decreases nonlinearly as a
function of the radius $a$, as given by the uniform formula \eqref{asympto2}. Figure
\ref{f:fig5} recovers the three regimes of \eqref{asympto2}: the uncrowded regime
for $a<0.2L$, where the effective diffusion coefficient does not show any
significant decrease, a region $0.2L<a<0.4L$, where the leading order term of
the effective diffusion coefficient is logarithmic, and for $a>0.4L$ the
effective diffusion coefficient decays as $\sqrt{(L-2a)/L}$, in agreement with
(\ref{asympto2}).

Finally, to estimate the density of obstacles in a neuron from \eqref{asympto2}, \eqref{lambda}, and \eqref{barD}, a reference density has to be chosen. The reference diffusion coefficient is chosen to be that of receptors moving on a free membrane (with removed cholesterol), estimated to be $0.17\leq D\leq0.2\,\mu$m$^2$/sec \cite{HolcmanSchuss2015}, while with removing actin, the diffusion coefficient is $0.19\,\mu$m$^2$/sec. The reference value $D=.2\mu$m$^2$/s gives an estimate of the crowding effect based on the measured diffusion coefficient (Fig. \ref{f:fig5}d). The reduction of the diffusion coefficient from $D=0.2\,\mu$m$^2$/sec to $D=0.04\,\mu$m$^2$/sec is achieved when 70\% of the membrane surface is occupied by obstacles. Thus obstacles impair the diffusion of receptors and are therefore responsible for the large decrease of the measured diffusion coefficient (up to 5 times).

%%%%%%%%%%%%%%%%%%%%%%%%%%%%%%%%%%%%%%%%%%%%%%%%%%%%%%%%%%%%%%%%%%%%%%%%%%%%%%%%%%%%%%
To conclude, as illustrated in fig. \ref{fig:drawing}, diffusion in a crowded membrane involves various time regimes: at very short time scale, before a Brownian particle has the time to escape between small aperture near obstacles, the particle diffuses freely, characterized by the homogeneous membrane diffusion coefficient. This approximation is valid before the NET time scale is reached (see formula \ref{bartau}). For longer time, $t \gg \bar\tau$, the motion of a Brownian particle is characterized again by a diffusion process, but now the diffusion coefficient does account for the obstacles and in the very density limit, the effective diffusion coefficient is given by
\beq
D_e=D \frac{L^2}{2\pi(L^2-\pi R^2)}\sqrt{\frac{L-2R}{R}}, \hbox{ for } R\leq L/2.
 \eeq
At an intermediate time regime between the two extreme cases described above, a stochastic particle is hopping from one square to another, characterized as anomalous diffusion (blue curves in fig. \ref{fig:drawing}).

%%%%%%%%%%%%%%%%%%%%%%%%%%%%%%%%%%%%%%%%%%5
\section{Jump processes for a model of telomere length dynamics}
%%%%%%%%%%%%%%%%%%%%%%%%%%%%%%%%%%%%%%%%%%
Stochastic jumps are inherent to physical and biological processes that can be studied in various limits (diffusion  approximation \cite{Schuss2}). We review here the example of a telomere (end of a chromosome) model introduced in \cite{Zhou,Daoduc}, in which the length $x$ of the telomere can decrease or increase at each division.

The length $x$ decreases by a fixed length $a$ with probability $l(x)$ or, if recognized by a polymerase, it increases by fixed length $b$ with probability $r(x)=1-l(x)$. The jump probability $r(x)$ is a decreasing function of $x$ with $r(0)=1$. Thus the length of the telomere at division $n$ is an asymmetric random walk $x(n)$. In this simplified model, the maximal length of a telomere is $L\gg b$. When the length falls below a critical value $T$, cell division stops.

%%%%%%%%%%%%%%%%%%%%%%%%%%%%%%%%
\subsection{The asymmetric random jump model}
%%%%%%%%%%%%%%%%%%%%%%%%%%%%%%%%
The model of the telomere dynamics is
\begin{align}
x_{n+1}=
\left\{\begin{array}{ll}x_n-a&\mbox{w.p.}\
l(x_n)\\ \\
x_n+b&\mbox{w.p.}\ r(x_n),\end{array}\right.\label{dynamics}
\end{align}
where the right-probability $r(x)$ can be approximated by
\begin{align}
r(x)=\frac{1}{1+\alpha x},\label{rx}
\end{align}
for some $\alpha>0$. Scaling $x_n=y_nL$ and setting $\eps=b/L$, the dynamics \ref{dynamics} becomes
\begin{align}
y_{n+1}=\left\{\begin{array}{ll}y_n- \ds\frac{\eps a}{b}&\mbox{w.p.}\
\tilde l(y_n)\\
y_n+\eps &\mbox{w.p.}\ \tilde
r(y_n),\end{array}\right.\label{tdynamics}
\end{align}
where $\tilde l(y)=l(x)$ and $\eps T/b<y<1$. In the limit $\eps\ll1$, the process $y_n$
moves in small steps. The dynamics (\ref{tdynamics}) falls under the general scheme (see \cite{Schuss2})
\begin{equation}
y_{n+1}=y_n+\eps \xi _n,  \label{2RW}
\end{equation}
where
\begin{align}
\Pr\left\{ \xi _n=\xi \,|\,y_n=y,\,y_{n-1}=y_1,\,\ldots\right\}
=w(\xi\,|\,y,\eps),\label{Pxi}
\end{align}
$\eps$ is a small parameter, and $y_0$ is a random variable \index{random variable} with a given pdf $p_0(y)$. In the case at hand the function $w(\xi\,|\,y)$ defined in (\ref{Pxi}) is given by
\begin{align}
w(\xi\,|\,y)=(1-\tilde
r(y))\delta\left(\xi+\frac{a}{b}\right)+\tilde r(y)\delta(\xi-1),\label{wxiy}
\end{align}
The pdf of $y_n$ satisfies %the forward master equation
%\begin{align}
%&\,p_{\eps}(y,n+1\,|\,x,m)\label{MA}\\
%=&\,\left\{\begin{array}{lll}
%p_{\eps}\left(y+\eps\ds\frac{a}{b},n\,\Big{|}\,x,m\right)\tilde
%l\left(y+\eps\ds\frac{a}{b}\right)+p_{\eps}(y-\eps,n\,|\,x,m)\tilde
%r(y-\eps)&\mbox{for}&\eps<y<1-\eps\ds\frac{a}{b}\\
%&&\\
%p_{\eps}\left(y+\eps\ds\frac{a}{b},n\,\Big{|}\,x,m\right)\tilde
%l\left(y+\eps\ds\frac{a}{b}\right)&\mbox{for}&0<y<\eps
%\end{array}\right.\nonumber
%\end{align}
%and
the backward equation for $0<x<1$
 \begin{align}
&\, p_{\eps}(y,n\,|\,x,m)-p_{\eps}(y,n\,|\,x,m+1)\label{2BKE}\\
=&\,p_{\eps}\left(y,n\,\Big{|}\,x-\eps\frac{a}{b},m+1\right)\tilde
l(x)+p_{\eps}(y,n\,|\,x+\eps,m+1)\tilde r(x)-p_{\eps}(y,n\,|\,x,m).\nonumber
 \end{align}
The first conditional jump moment, $m_1(y)=- a\tilde l(y)/b+\tilde r(y)$ changes sign
at
$z_0=b/L\alpha a$, so $p_{\eps}(y,n)$ converges to a quasi-stationary density
$p_{\eps}(y)$
for large $n$, before the trajectory $y_n$ is terminated at $y=T/L$.

One dimensional processes (see eq. \ref{2RW}) in the small jump limit is described in \cite{Schuss2} p.236 and p.303, see also \cite{knessl:1985bs,matkowsky:1984,knessl:1984ij}. In this limit, the Kramers-Moyal approximation consists in expanding in $\eps$ and then approximating equation \ref{2BKE} by a direct truncation to a second order equation. The method is equivalent to construct a stochastic processes with the first and second moments that match the coefficients of the  Kramers-Moyal approximation. We present below the WKB construction of an approximated solution.

%The quasi-stationary master equation is
%\begin{align}
%p_{\eps}(y)=p_{\eps}\left(y+\eps\frac{a}{b}\right)\tilde
%l\left(y+\eps\frac{a}{b}\right)+p_{\eps}\left(y-\eps \right)\tilde
%r\left(y-\eps\right)\hspace{0.5em}\mbox{for}\ \eps<y<1-\eps\frac{a}{b},\label{stMA}
%\end{align}
%which for small $\eps$ is peaked near $z_0$ (see \cite{holcmanschusstelomere}). Because the process $y_n$ cannot reach a point $y\leq\eps$ by jumping to the right, equation \eqref{stMA} reduces for $y\leq\eps$ to
%\begin{align}
%p_{\eps}(y)=p_{\eps}\left(y+\eps\frac{a}{b}\right)\tilde
%l\left(y+\eps\frac{a}{b}\right). \label{stMAr}
%\end{align}
%%%%%%%%%%%%%%%%%%%%%%%%%%%%%%%%%%%%%%%%%%%%%%%%%%%%
\subsection{Construction of the quasi steady-state density $p_{\eps}(y)$}
%%%%%%%%%%%%%%%%%%%%%%%%%%%%%%%%%%%%%%%%%%%%
The structure of the quasi-stationary solution $p_{\eps}(y)$ for $\eps\ll1$ and $y>\eps$, can be obtained in the WKB form \cite{Schuss2}
\begin{align}
p_{\eps}(y)=K_\eps(y)\exp\left\{-\frac{\psi(y)}{\eps}\right\},\label{WKBsltn}
\end{align}
where
\begin{align}
K_\eps(y)=\sum_{i=0}^\infty\eps^iK_i(y)
\end{align}
and $K_i(y)$ are chosen such that $p_{\eps}(y)$ is normalized. The components of the expansion (\ref{WKBsltn}) are found by  solving the eikonal equation for $\psi(y)$
 \begin{align}
  \int\limits_{\rR}\left[e^{\xi\psi'(y)}-1\right]w(\xi\,|\,y,0)\,d\xi=0,\label{2eikonal}
 \end{align}
which for the problem at hand takes the form
\begin{align}
\tilde l(y)\exp\left\{-\psi'(y)\frac{a}{b}\right\}+\tilde
r(y)\exp\left\{\psi'(y)\right\}=1\hspace{0.5em}\mbox{for}\ y>\eps.\label{eikonaleq}
\end{align}
%The derivatives of $\psi(y)$ at $z_0$ can be computed by expanding (\ref{2eikonal})
%near $y=z_0$ in powers of $\xi\psi'(y)$ to obtain
%\begin{align}
%\psi''(z_0)=-\frac{2m_1'(z_0)}{m_2(z_0)}.\label{psi''}
%\end{align}
%First, we consider the eikonal equation (\ref{eikonaleq}) near $z_0$, where $\psi'(z_0)=0$, and find the local expansion of $\psi(y)$ in powers of $y-z_0$. The expansion gives
%\begin{align} \label{psi}
%\psi''(z_0)=\frac{2aL\alpha}{a+b},\quad
%\psi'''(z_0)=-\frac43\frac{(a+2b)\alpha^2L^2a^2}{b(a+b)^2},
%\end{align}
%and so on. It is known \cite{DSP} that $\psi(y)$ is convex, has a minimum at $z_0$, and can be assumed to vanish at the minimum, that is, $\psi(z_0)=0$ .

At the next order in $\eps$, we find that $K_0(y)$ is the solution of the ``transport" equation
\begin{align}
\int\limits_{\rR}\left\{\frac{\p}{\p y}[w(\xi\,|\,y)K_0(y)]+\frac{\xi w(\xi
\,|\,y)}{2}\psi''(y)K_0(y)\right\}\xi
 e^{\xi\psi'(y)}\,d\xi=0.\label{2transport}
\end{align}
Equation \eqref{2transport} has a removable singularity at $y=z_0$, because the coefficient of $K_{0y}$ in \eqref{2transport},
\begin{align*}
\int\limits_{\rR}\xi w(\xi\,|\,y)e^{\xi\psi'(y)}\,d\xi=&\,\int\limits_{\rR}\xi
w(\xi\,|\,y)\left[1+\xi\psi''(z_0)(y-z_0)+O\left((y-z_0)^2\right)\right]
\,d\xi\\
=&\,[m_1'(z_0)+m_2(z_0)\psi''(z_0)](y-z_0)+O\left((y-z_0)^2\right)\\
=&\,-m_1'(z_0)(y-z_0)+O\left((y-z_0)^2\right),
\end{align*}
vanishes linearly at $z_0$. However, the coefficient of $K_0(y)$ in \eqref{2transport} also vanishes at
$z_0$.

%%%%%%%%%%%%%%%%%%%%%%%%%%%%%%%%%%%%%%%%%%%%%%%%%%%%
\subsection{Extreme statistics for the shortest telomere}
%%%%%%%%%%%%%%%%%%%%%%%%%%%%%%%%%%%%%%%%%%%%
In this section, we present a different approach for computing the steady solution of the pdf for the process \ref{dynamics}. We derive the Takacs equation and then study the statistics of the shortest trajectory (shortest telomere) for an ensemble of n identical independently distributed (iid) processes.

The steady state distribution associated to the telomere equation \ref{dynamics} can be rescaled with a constant drift for shortening, and possible large jumps with exponential rates for elongation. The jump rate function becomes $\lambda(X)=\dfrac{1}{1+B X}$, and the probability for the jump $\bar{\xi}$ is given by $Pr(\bar{\xi}=y)=\theta e^{-\theta y} = b(y)$, where $B= a \beta$ and $\theta= ap$. The pdf $f(x,t) = \partial F(x,t)/\partial x$ where  $F(x,t) = \text{Pr}\left\{ X(t) \leq x \right\}$ satisfies the Takacs equation, which is written for the forward Fokker-Planck equation $x>0$,
\beq
 \partial_t f =\partial_x f - \lambda(x)f(x,t) + \int_{0}^l\lambda (y)f(y,t)b(x-y)dy.
 \label{Meq}
\eeq
The stationary distribution function is \cite{Daoduc}
 \beq
\bar{f}(x) &=& \dfrac{\theta\left[ \theta( x+\frac{1}{B}) \right]^{\frac{1}{B}}e^{-\theta\left(x+\frac{1}{B}\right)}}{\Gamma \left( \frac{1}{B}+1,\frac{\theta}{B}\right)},
\label{stationarydistrib}
\eeq
where $\Gamma (s,x) = \int_{x}^{+\infty} t^{s-1}e^{-t}dt$ is the upper incomplete Gamma function. Now the distribution of the shortest telomere \cite{Daoduc} in an ensemble of $2n$ telomeres, corresponding to a total of $n$ chromosomes (16 in yeast and $n$ is in the range of $36-60$) is estimated when their lengths are independent identically distributed variables $L_1,L_2 , \ldots L_{2n}$. Considering $2n$ iid variables $X_1, \ldots X_{2n}$ following a distribution $f$, the pdf of the minimum $X_{(1:2n)}=\min (X_1,X_2 , \ldots X_{2n})$ is given by
\beq\label{shortestpdf}
f_{X_{(1:2n)}}(x) = 2n(1-F)^{2n-1}(x)f(x),
\eeq
where $F(x)=\int_0^{x} f(u)du$. The statistical moments $ \bar{X}^k_{(1:2n)}=  \int_\mathbb{R+}x^{k}f_{X_{(1:2n)}}(x)dx$
are given by
\beq
\bar{X}^k_{(1:2n)}=  k\int_\mathbb{R+}x^{k-1}(1-{F})^{2n}(x)dx.\label{shortestmom}
\eeq
When $f$ is a Gamma distribution of parameter $\alpha$ and $n$ sufficiently large, $\bar{X}^k_{(1:2n)}$ can be estimated using the Laplace's method.

In the limit $x$ tends to 0,  Eq.\eqref{stationarydistrib} with $\alpha = 1 + \frac{1}{B}$ satisfies $F(x) \approx m x^r$  with $m=\frac1{\alpha\Gamma(\alpha)}>0$ and $r=\alpha>1$,
\beq\label{laplace}
 \bar{X}^k_{(1:2n)} =
 \frac{k}{r}\int_0^{+\infty} x^{k/r-1} \exp[2n\ln(1- F(x^{1/r}))]dx \approx
\bar{X}^k_{(1:2n)} \approx \frac{k\Gamma\left(\frac{k}{\alpha} \right)}{\alpha}\left(\frac{\alpha \Gamma(\alpha)}{2n} \right)^{k/\alpha}.
\eeq
Using formulae \eqref{shortestpdf}, \eqref{shortestmom} and \eqref{laplace}, the pdf and the moments of the shortest telomere length $L_{1:2n}$ for $k=1$ can be estimated and the shortest telomere length is
 \beq
 \bar{L}_{(1:2n)} \approx L_0-\frac{1}{B}+ \dfrac{\Gamma\left(1+\frac{1}{B}\right)\Gamma\left(\frac{1}{1+1/B}\right)^{\frac{1}{1+1/B}}}{p\left(1+\frac{1}{B}\right)^{\frac{1}{1+B}}(2n)^{\frac{1}{1+1/B}}}\label{telshort1}.
 \eeq
Using the values $p=0.026$, $\beta=0.045$ ($B = 0.16 <0.5$) and $L_0=90$ (yeast) and eq. \eqref{shortestmom} for k=1 and 2, the mean shortest telomere length is $184 \pm 25$ bps.

To estimate the gap between the shortest telomere and the others, we shall compute the distribution of the second shortest length $X_{(2:2n)}$. The pdf $f_{X_{(2:2n)}}$ of $X_{(2:2n)}$ is given by
\beq
f_{X_{(2:2n)}}(x) = 2n(2n-1)F(x)(1-F(x))^{2n-2}f(x), \label{second}
\eeq
and the statistical moments $\bar{X}^k_{(2:2n)}$ satisfy the induction relation
\beq
\bar{X}^k_{(2:2n)} = n\bar{X}^k_{(1:2n-1)} -(2n-1) \bar{X}^k_{(1:2n)}.
\label{X2}
\eeq
Using equation \eqref{laplace} for $k=1$, we obtain that the ratio $\frac{\bar{X}_{(2:2n)}}{\bar{X}_{(1:2n)}}$, for $n$ or $B >> 1$  is given asymptotically for $\alpha = 1+ \frac{1}{B}$ by
 \beq
\frac{\bar{X}_{(2:2n)}}{\bar{X}_{(1:2n)}} \approx 1+ \frac{1}{\alpha} \label{ratio}.
\eeq
To conclude, for a pdf with a nonzero first order derivative at 0, this ratio is a universal number $\frac{3}{2}$. In the case of yeast, eq.\eqref{X2} reveals that the mean length of the second shortest telomere is 207 bps. Thus, the shortest telomere is on average 22 bps shorter than the second one. This gap results from the statistical property of the telomere number and dynamics and should exist in all species. It suggests that the length of the shortest telomere controls the number of division. The computation of the mean time to threshold is more involved as it requires finding an approximation of the pdf in two time intervals. Interestingly, this time depend on the escape of a coarse-grained stochastic dynamics from an effective potential \cite{HStelomere}.

%%%%%%%%%%%%%%%%%%%%%%%%%%%%%%%%%%%%%%%%%%%%%%%%%%%%%%
\section{Hybrid discrete-continuum modeling for stochastic gene expression within a autoregulatory positive feedback loop}
%%%%%%%%%%%%%%%%%%%%%%%%%%%%%%%%%%%%%%%%%%%%%%%%%%%%%%
We end this tribute to Smoluchowski by a description of stochastic modeling of gene activation and regulation. The difficulty in such modeling is the presence of the continuum and discrete description to account for few mRNA (discrete) and large synthesized proteins (continuum). Gene expression is often model by classical Mass-Action laws with additive noise. We present here an alternative approach based on Markov jump processes and we the large number approximation to simplify the equation. The result is a hybrid continuum-discrete ensemble of equations that can give different predictions than classical model.  The model is applied to a positive feedback loop of gene regulation based on a transcription factor called Krox20 \cite{Bouchoucha}.

We recall that proteins are produced by mRNAs. Sometimes such mechanism involves a feedback control of the proteins on the gene to regulate the mRNA production. For a positive feedback, the produced proteins bind the gene sites to activate the mRNA production. A transcription factor such as Krox20 positively regulates its own expression and results in a bistable switch: either proteins are expressed or not.  The model is also used to extract parameters from data and predict the level of expression inside a cell population. Krox20 is involved in the hindbrain anterior-posterior identity, where 7-8 segments called rhombomeres are formed and the transcription factor is required for the particular construction of rhombomeres 3 and 5. The stochastic model of Krox20 expression is based on the interaction between mRNA and proteins, cooperative binding/unbinding of Krox20 proteins to four binding sites on the DNA called $A$ (Fig. \ref{Loop}). The difficulty in such a model is that only a few mRNA molecules are involved in the activation process, which is coupled to a continuous description of  proteins.
%%

%%%%%%%%%%%%%%%%%%%%%%%%%%%%%%%%%%%%%%%%%%%%%%%%%%%%%%
\subsection{Stochastic model of gene activation}
%%%%%%%%%%%%%%%%%%%%%%%%%%%%%%%%%%%%%%%%%%%%%%%%%%%%%%
To compute the number of proteins, we need to follow simultaneously three variables: the state $s$ of element A, the number $m$ of {\it Krox20} mRNA and the number $n$ of unbound Krox20 proteins. The joint probability $p_s(m,n,t)$ to find element A in state $s$ with $m$ {\it Krox20} mRNA molecules and $n$ free Krox20 proteins is
\beq
p_s(m,n,t)=Pr\{s(t)=s, mRNA(t)=m, Krox20(t)=n\}
\eeq
and it satisfies a Master equation \cite{Schuss2}, where only one binding or unbinding occurs at a time
\beq
\frac{\p p_s}{\p t}(m,n,t) =   \Phi_s  p_s(m-1,n,t) +  (m+1) \Psi p_s (m+1,n,t) -(\Phi_s + m\Psi) p_s(m,n,t)  \nn \\
 + m \phi p_s(m,n-1,t) +  (n+1) \psi p_s (m,n+1,t) - ( m\phi + n\psi) p_s(m,n,t) \nn \\
   + \mu_{s+1} p_{s+1}(m,n-1,t) +  (n+1) \lambda_{s-1}  p_{s-1}(m,n+1,t) - ( \mu_s + n \lambda_s)  p_s(m,n,t). \vspace{-3cm}  \label{MasterEquation0}  %\\
\eeq
The mRNA production rate is given function
\beq
\Phi_s(t) = \Phi_I\theta(t_I-t) +  \Phi_{A,s}
 \eeq
 where $\theta(t)$ is the Heaviside function. There function models an initial phase where proteins are produced until time $t_I$. During this phase, mRNA is produced with a Poissonian rate $\Phi_I$ by an external molecule. Each mRNA protein is degraded with a Poissonian rate $\Psi$. Proteins are produced with a Poissonian rate $\phi$ and are degraded with a rate $\psi$ and can bind to a promoter site that has $N_b=4$ binding sites. The state of A is characterized by s=0, 1, 2, 3, 4  bound molecules. The autoregulatory production of mRNA occurs with Poissonian rates $\Phi_{A,s}= \Phi_A \xi_s$ that depend on state $s$, where $\Phi_A$ is the maximal production rate and $\xi_s$ describes the modulation due to the state of element A. Binding and unbinding of proteins to A are described by state dependent binding and unbinding rates $\lambda_s$ and $\mu_s$.

When the change in the number of free Krox20 proteins $n$ due to binding and unbinding to element A is neglected, in the limit $n\gg N_b$,  the Master eq.~\ref{MasterEquation0} can be approximated by
\beq
\frac{\p p_s}{\p t}(m,n,t) =
\Phi_s  p_s(m-1,n,t) +  (m+1) \Psi p_s (m+1,n,t) -(\Phi_s + m\Psi) p_s(m,n,t)  \nn \\
+ m \phi p_s(m,n-1,t) +  (n+1) \psi p_s (m,n+1,t) - ( m\phi + n\psi) p_s(m,n,t) \nn \\
+ \mu_{s+1} p_{s+1}(m,n,t) +  n \lambda_{s-1}  p_{s-1}(m,n,t) - ( \mu_s + n \lambda_s)  p_s(m,n,t)  \label{MasterEquation}
\eeq
The first line in eq.~\ref{MasterEquation} describes the production and degradation of a mRNA, while the second is for the production and degradation of a Krox20 protein. The last one is for the binding and unbinding of a Krox20 protein to element A. The marginal probabilities is
\beq\label{jointProb}
p_s(m,n,t) = \sum_{s'=0}^{N_b} p_{s,s'}(m,n,t)= \sum_{s'=0}^{N_b} p_{s',s}(m,n,t)\,.
\eeq
Binding and unbinding to element $A$ is fast compared to the turn over of proteins, thus we use the approximation
\beq
p_s(m,n,t)\approx p(m,n,t)p_s(n),
\eeq
where $p_s(n)$ are the steady-state probabilities to find element A in state $s$ for a given number of Krox20 proteins $n$, and $p(m,n,t)$ is the probability to find $m$ mRNA molecules and $n$ proteins at time $t$.

The steady state condition for binding and unbinding from eq.~\ref{MasterEquation} is
\beq
\mu_{s+1} p_{s+1}(n) = n \lambda_s  p_s(n),
\eeq
leading to the solution
\beq\label{steadystateprob}
p_s(n) =  \frac{ \prod_{j=0}^{s-1} n \gamma_j}{\sum_{i=0}^{N_b} \prod_{j=0}^{i-1} n \gamma_j }\,,
\eeq
where
\beq\label{defgammai}
\gamma_i =\frac{\lambda_i}{\mu_{i+1}}\,,\quad i=0,\ldots N_b-1\,.
\eeq
The effective mRNA production rates is defined by
\beq\label{defPhi_A(n)}
\Phi_A(n) =  \sum_{s=0}^{N_b} \Phi_{A,s} p_s(n) = \Phi_A \sum_{s=0}^{N_b} \xi_s p_s(n)\,.
\eeq
At this stage, the approximated Master equation for the joint pdf $p(m,n,t)$ is
\beq\label{MasterEquation_ss}
\frac{\p p_s}{\p t}(m,n,t) =  &=&   \Phi(n,t)  p(m-1,n,t) +  (m+1) \Psi p(m+1,n,t) -(\Phi(n,t) + m\Psi) p(m,n,t)  \nn \\
&& + m \phi p(m,n-1,t) +  (n+1) \psi p(m,n+1,t) - ( m\phi + n\psi) p(m,n,t)
\eeq
The marginal probabilities for mRNA molecules, Krox20 proteins and the state of element A are
\beq\label{marginalProb}
\begin{array}{rcl}
\ds p(m,t) &=&  \ds \sum_{s=0}^{N_b} \sum_{n=0}^{\infty} p_s(m,n,t) =  \sum_{n=0}^{\infty}  p(m,n,t)\,, \\
\ds p(n,t) &=& \ds \sum_{s=0}^{N_b} \sum_{m=0}^{\infty} p_s(m,n,t)= \sum_{m=0}^{\infty} p(m,n,t) \,, \\
\ds p_s(t) &=& \ds  \sum_{m=0}^{\infty} \sum_{n=0}^{\infty}p_s(m,n,t) =  \sum_{n=0}^{\infty}   p(n,t)  p_s(n)\,.
\end{array}
\eeq
It remains difficult to study equations \ref{MasterEquation_ss} because  $n$ can be large while m is of the order of few and thus there is no clear limit approximations.  The long-time asymptotic is however an important quantity to estimate and in particular how it depends on initial conditions. This limit tells us whether or not a cell expresses Krox20. The diffusion approximation leads to a dynamical system that predicts a two state attractors characterized by full or zero expression, while the Master system shows that there is be a continuum level of expression, but the distribution is dominated by two peaks. This difference justifies the need of a Markov chain description, in particular to study cells at the boundary between regions, where a graded expression is predicted.
%
%%
%%%%%%%%%%%%%%%%%%%%%%%%%%%%%%%%%%%%%%%%%%%%%%
\subsection{Mean-field approximation}
%%%%%%%%%%%%%%%%%%%%%%%%%%%%%%%%%%%%%%%%%%%%%
The mean field approximation for the mean quantities is based on the scaled variables
\beq
\tau = \Psi t\,,  \quad \hat m = \frac{m}{m_0}\,, \quad \hat n = \frac{n}{n_0}\
\eeq
and the normalized parameters are
\beq\label{scaledparam}
\ds m_0= \frac{\Phi_A }{\Psi} \,,\quad n_0=m_0\frac{\phi}{\psi} \,,\quad \alpha = \beta n_0 \,, \quad \epsilon = \frac{\Psi}{\psi}\,,\quad \chi =\frac{\Phi_I}{\Phi_A},
\eeq
where the average values $m_0$ and $n_0$ characterize the mRNA and proteins when element A is fully activated. The mean-field equation in the scaled variables $\hat m$ and $\hat n$ is expressed using the function
\beq\label{deffuncf}
f(x) =  \sum_{s=0}^{N_b} \xi_s  \frac{ \prod_{j=0}^{s-1} x \tilde \gamma_j}{\sum_{i=0}^{N_b} \prod_{j=0}^{i-1} x \tilde \gamma_j }.
\eeq
Indeed, the Kramers-Moyal expansion of eq.~\ref{MasterEquation_ss} is
\beq
\frac{\p p(\hat m,\hat n,\tau)}{\p \tau}
 &=&  \sum_{k=1}^\infty  \( \frac{1}{m_0}\)^{k-1}\frac{\p_{\hat m}^k}{k!}  \( (-1)^k (\chi \theta(\tau_I-\tau) +  f(\alpha \hat n)) +   \hat m \)   p(\hat m,\hat n,\tau)  \nn\\
 && + \frac{1}{\epsilon} \sum_{k=1}^\infty \( \frac{1}{n_0}\)^{k-1}\frac{\p_{\hat n}^k}{k!}  \( (-1)^k  \hat m + \hat n \)  p(\hat m,\hat n,\tau)
\eeq
From truncating the series at first order and neglecting the initiation ($\chi=0$), we obtain the first order dynamical system, which is the mean-field equation \cite{Schuss2}:
\beq\label{dynamicalsystem}
\begin{array}{rcl}
\ds\frac{d}{d\tau} \hat m &=& \ds \chi \theta(\tau_I-\tau) + f(\alpha \hat n) -  \hat m \\ &&\\
\ds \frac{d}{d\tau} \hat n &=& \ds \frac{1}{\epsilon} \( \hat m -\hat n \)\,.
\end{array}
\eeq
The fixed points are given by $\hat n^*=\hat m^*$ in eq.~\ref{dynamicalsystem}. When $\chi=0$, there is a minimal value $\alpha_{min}\approx 33.8$ for which $\alpha < \alpha_{min}$, there is a single fixed point $\hat m^*=0$ ($\alpha_{min}$, computed using the solution $z_0$ of $\frac{f'(x)}{f(x)}=\frac{1}{x}$. The minimum value is $\alpha_{min} = \frac{z_0}{f(z_0)}= \frac{1}{f'(z_0)}$).

For $\alpha >\alpha_{min}$, there are two stable fixed points $\hat m^*_{1}$ and $\hat m^*_{3}$ and an unstable saddle point $\hat m^*_{2}$ (Fig.~\ref{results} Upper). For large $\alpha$, the asymptotic values are $\hat  m_2^*\to 0$ and $\hat m_3^*\to 1$. Th value $\hat m_3^* \approx 1$ ($m_3^*=m_0\hat m_3^* \approx \frac{\phi_A}{\Psi} $) corresponds to the situation where element A is fully activated (Fig.~\ref{results}).

The two stable fixed points defines two basins of attraction and a saddle point for $\alpha > \alpha_{min}$, contained in a separatrix (Fig.~\ref{results} Upper for $\alpha=40$). For an initial conditions outside the basin of attraction of the fixed point $\hat m_1^*$, the dynamics evolves towards the Up attractor (a high protein expression level). In contrast, for a small amplitude initiation, the expression vanishes. In Fig.~\ref{results} (Upper left) shows the separatrix for different values of $\alpha$: with increasing $\alpha$ the basin of attraction of $\hat m_1^*=0$ shrinks and asymptotically vanishes for large $\alpha$. In summary, the mean field dynamical system underlying protein activation shows a bistable behavior between two attractors depending on the initial condition, which can be seen as a random variable
%
%%%%
%\subsection{Analysis of the stochastic model}
%%%%%
In contrast, the numerical analysis of the Master equation \ref{MasterEquation_ss} for the probability $p(m,n,t)$ with the scaled variables $\hat m$ and $\hat n$, $p(\hat m,\hat n,t)$ for $p(m,n,t)$ uses the marginal probabilities defined in eq.~\ref{marginalProb}. The mean values for $\hat m$ and $\hat n$ are given by
\beq
\begin{array}{rcl}
\ds \overline{\hat m}(t) &=& \ds \frac{1}{m_0}   \sum_{n=0}^{\infty}  m p(m,n,t) = \ds \frac{1}{m_0}   \sum_{m=0}^{\infty}  m p(m,t) =\frac{\bar m (t)}{m_0} \,,\\
\ds \overline{\hat n}(t) &=& \ds \frac{1}{n_0}   \sum_{m,n=0}^{\infty}  n p(m,n,t)= \ds \frac{1}{n_0}   \sum_{n=0}^{\infty}  n p(n,t)=\frac{\bar n (t)}{n_0}
\end{array}
\eeq
In \cite{Bouchoucha}, the analysis reveals that for $\beta > 0.13$ the probability distribution $p(\hat n)$ at time $t=500\min$ is bimodal with peaks at zero and close to one (Fig.~\ref{results} Lower panel).

To conclude, simulating the Master equations reveals a continuum of steady state characterized by two peaks, while the mean field approximation predict a bistable distribution, where the dynamics can fall into one of two attractors. More stochastic analysis is expected to reveal in the future how gene expression regulate development, boundary between brain regions \cite{reingruberholcman2015} or diseases.
%%%%%%%%%%%%%%%%%%%%%%%%%%%%%%%%%%%%%%%%%%%%%%%%%%%%%%%%%%%%%%%%%%%%%%%%%5
\section{General conclusion and perspective}
%%%%%%%%%%%%%%%%%%%%%%%%%%%%%%%%%%%%%%%%%%%%%%%%%%%%%%%%%%%%%%%%%%%%%%%%%5
We reviewed here the influential Smoluchowski equation and its applications in modeling, analysis in biophysics and computational cell biology. In general stochastic processes have now become the framework for extracting features from molecular and cellular large data sets. In that context, the Narrow Escape Theory is a coarse-graining procedure  revealing how structures (geometry) controls physiology through time scales and rare random events. The Smoluchowski equation is also the basis for analysis super-resolution data and obtained deconvolution algorithms for extracting biophysical parameters \cite{holcmanBJ2015}. The analysis of diffusion with obstacles reveals how narrow passage between obstacles defines the effective measured diffusion.

Polymer models allow computing looping rates used to interpret large data about the position of chromosomes inside the cell nucleus.  We also illustrated stochastic processes in system biology by presenting a stochastic model that couples continuum and discrete levels. Stochastic gene activation, mRNA and protein productions remain an exciting field where the model analysis remains difficult due to the large degree of freedom (large parameter space). A general framework to extract parameters and study feedback loop in gene activation is still to be found.

Although stochastic chemical reactions are now routinely simulated using the classical Gillespie's algorithm, exploring the parameter space can be done when possible using asymptotic formulas, derived from the model equations.  Another area that we have not reviewed here is the recent development of aggregation-dissociation with a finite number of particles. Smoluchowki fragmentation-aggregation model is an infinite set of equations that described colloids in solution and other molecular aggregations. However, aggregation with a finite number of particles to study viral capsid formation and telomere dynamics in the nucleus requires a different probability framework than infinite set of equations \cite{hoze2013,hozeJMB2015}, leading to novel quantity to estimate such as the time spent by two particles in the same cluster.

%%%%%%%%%%%%%%%%%%%%%%%%%%%%%%
\end{document}